\documentclass[11pt,a4paper]{article}
\usepackage{cite}

\usepackage{amsmath, amsthm,mathtools, url,amssymb,upgreek,slashed}
\usepackage{ifpdf}
\ifpdf
  \usepackage[pdftex]{graphicx}
  \usepackage{epstopdf}
\else
  \usepackage[dvips]{graphicx}
\fi
\parskip=\baselineskip

\begin{document}
\begin{titlepage}
\begin{flushright}

\end{flushright}

\vskip 1.5in
\begin{center}
{\bf\Large{Matrix Models and Deformations of JT Gravity}}
\vskip
0.5cm { Edward Witten} \vskip 0.05in {\small{ \textit{Institute for Advanced Study}\vskip -.4cm
{\textit{Einstein Drive, Princeton, NJ 08540 USA}}}
}
\end{center}
\vskip 0.5in
\baselineskip 16pt
\begin{abstract}  Recently, it has been found that JT gravity, which is a two-dimensional theory
with bulk action $ -\frac{1}{2}\int {\mathrm d}^2x \sqrt g\phi(R+2)$, is dual to a matrix model, that is, a random ensemble of quantum
systems rather than  a specific quantum mechanical system.   In this article, we argue that
a deformation of JT gravity with bulk action $ -\frac{1}{2}\int {\mathrm d}^2x \sqrt g(\phi R+W(\phi))$ is likewise
dual to a matrix model.   With a specific procedure for defining the path integral of the theory,
we determine the density of eigenvalues of the dual matrix model.   There is a simple answer if $W(0)=0$, and otherwise
a rather complicated answer.  
   \end{abstract}
\date{May, 2020}
\end{titlepage}
\def\SO{{\mathrm{SO}}}
\def\G{{\text{\sf G}}}
\def\la{\langle}
\def\Erf{{\mathrm{Erf}}}
\def\sC{{\mathrm C}}
\def\ra{\rangle}
\def\g{{\cmmib g}}
\def\U{{\mathrm U}}
\def\M{{\mathcal M}}
\def\d{{\mathrm d}}
\def\CS{{\mathrm{CS}}}
\def\Z{{\Bbb Z}}
\def\R{{\Bbb R}}
\def\J{{\mathcal J}}
\def\Bbb{\mathbb}
\def\Tr{{\rm Tr}}
\def\diff{{\mathrm{diff}}}
\def\SS{{\sf S}}
\def\16{{\bf 16}}
\def\1{{(1)}}
\def\2{{(2)}}
\def\3{{\bf 3}}
\def\4{{\bf 4}}
\def\Sch{{\mathrm{Sch}}}
\def\b{{\mathrm b}}
\def\sg{{\mathrm g}}
\def\GHY{{\mathrm{GHY}}}
\def\i{{\mathrm i}}
\def\h{\widehat}
\def\u{u}
\def\D{D}
\def\sp{{\sigma}}
\def\E{{\mathcal E}}
\def\O{{\mathcal O}}
\def\PF{{\mathit{P}\negthinspace\mathit{F}}}
\def\tr{{\mathrm{tr}}}
\def\be{\begin{equation}}
\def\ee{\end{equation}}
 \def\Sp{{\mathrm{Sp}}}
 \def\Spin{{\mathrm{Spin}}}
 \def\SL{{\mathrm{SL}}}
 \def\SU{{\mathrm{SU}}}
 \def\SO{{\mathrm{SO}}}
 \def\ll{\langle\langle}
\def\rr{\rangle\rangle}
\def\la{\langle}
\def\AdS{{\mathrm{AdS}}}
\def\ra{\rangle}
\def\T{{\mathcal T}}
\def\V{{\mathcal V}}
\def\bar{\overline}
\def\v{v}\
\def\SL{{\mathrm{SL}}}
\def\PSL{{\mathrm{PSL}}}
\def\kveps{\varepsilon}
\def\tilde{\widetilde}
\def\t{\widetilde}
\def\R{{\Bbb{R}}}\def\Z{{\Bbb{Z}}}
\def\N{{\mathcal N}}
\def\B{{\mathcal B}}
\def\H{{\mathcal H}}
\def\hat{\widehat}
\def\Pf{{\mathrm{Pf}}}
\def\PSL{{\mathrm{PSL}}}
\def\Im{{\mathrm{Im}}}

\numberwithin{equation}{section}

\def\neg{\negthinspace}
\def\d{\mathrm d}
\def\C{{\Bbb C}}
\def\HH{{\mathbb H}}
\def\P{{\mathcal P}}
\def\NS{{\sf{NS}}}
\def\Ra{{\sf{R}}}
\def\sV{{\sf V}}
\def\Z{{\Bbb Z}}
\def\A{{\eusm A}}
\def\B{{\eusm B}}
\def\S{{\mathcal S}}
\def\bar{\overline}
\def\sc{{\mathrm{sc}}}
\def\Max{{\mathrm{Max}}}
\def\CS{{\mathrm{CS}}}
\def\ga{\gamma}
\def\bg{\bar\ga}
\def\W{{\mathcal W}}
\def\M{{\mathcal M}}
\def\bM{{\overline \M}}
\def\L{{\mathcal L}}
\def\sM{{\sf M}}
\def\JT{{\mathrm{JT}}}
\def\gst{\mathrm{g}_{\mathrm{st}}}
\def\gstt{\widetilde{\mathrm{g}}_{\mathrm{st}}}
\def\veps{\varepsilon}

\def\be{\begin{equation}}
\def\ee{\end{equation}}

\tableofcontents

\section{Introduction and Discussion}\label{intro}

JT gravity is a simple model of a real scalar field $\phi$ coupled to gravity in two dimensions \cite{J,T}.   
For the case of negative cosmological constant,
the  bulk action in Euclidean signature is
\be\label{JTaction}I_\JT=  -\frac{1}{2}\int \d^2x \sqrt g\phi(R+2). \ee
Here $R$ is the scalar curvature of the metric tensor $g$. Usually one also adds  an Einstein-Hilbert action,
which in two dimensions is a multiple of the Euler characteristic, and also a Gibbons-Hawking-York boundary term.
It has been understood  recently \cite{SSS}  that this model is dual not to an ordinary quantum system on the asymptotic 
boundary of spacetime -- as one might have expected from prior experience with holographic duality -- 
but to a matrix model, a random ensemble of quantum mechanical systems.

The particular action in (\ref{JTaction}) was chosen to make a particularly simple model.  What happens
if we perturb away from this particular action?   It is natural to expect that there will still be a description by a random matrix
ensemble, presumably with a different distribution of eigenvalues.   This question will be investigated in the present paper.

Perturbations of JT with at most two derivatives are highly constrained.
Up to a Weyl transformation and a reparametrization of the scalar field $\phi$, one can assume \cite{Banks,GGG,Ikeda}
that the bulk action has the form 
\be\label{redto}I_W=-\frac{1}{2}\int\d^2x \sqrt g\left(\phi R + W(\phi)\right), \ee
with some function $W(\phi)$.   In JT gravity, $W(\phi)=2\phi$, and the dual quantum mechanical 
system (which in the case of JT gravity is really a random ensemble of quantum systems rather than a particular quantum system)
is defined in the asymptotic region
$\phi\to +\infty$.   We usually set $W(\phi)=2\phi + U(\phi)$, where $U(\phi)$ is the perturbation away from JT gravity.

The goal of this paper is to show, roughly speaking, that for an arbitrary $W$, this model is dual to a hermitian matrix model,
and to determine the eigenvalue distribution in the dual model.   By a hermitian matrix model, we mean a model of the same
class considered in \cite{SSS} (and in a great deal of earlier  work on two-dimensional gravity \cite{Wein,Kaz,David,Amb,KKM,BK,DS,GM,GMo,FGZ}).   These are double-scaled versions of a model of an $N\times N$  hermitian matrix $H$ with measure
$\exp(-N\,\Tr\,F(H))$ for some function $F(H)$ \cite{BPIZ}.    (In this application, $H$ is interpreted as the Hamiltonian of a dual quantum 
mechanical system.)   
Such a model is characterized by a function $\rho(E)$ which is the large $N$ limit
 of the density of eigenvalues of the matrix $H$.  Other observables
can be readily computed once $\rho(E)$ is known.    In a dual gravitational
theory, $\rho(E)$ is found by expressing the disc partition function $Z_D(\beta)$, where $\beta$ is the renormalized circumference
of the boundary of the disc,  as an integral over energies:
\be\label{injo} Z_D(\beta)=\int_{E_0}^\infty \d E \,\rho(E)\exp(-\beta E). \ee
Here $E_0$ is the threshold energy.   

With some caveats to be mentioned shortly, we will find a simple result for $\rho(E)$ if $W$ obeys one condition, namely $W(0)=0$.
This condition leads to $E_0=0$ -- the same value as in JT gravity -- and to the density of states
\be\label{onojjo} \rho(E;U)=e^{\SS_0}\left(\frac{\sinh 2\pi\sqrt E}{4\pi^2}+
\frac{e^{2\pi\sqrt E}U(\sqrt E)+e^{-2\pi\sqrt E}U(-\sqrt E)}{8\pi\sqrt E}\right),\ee
where $\SS_0$ is the ground state entropy.
It is also possible to get an exact formula for $\rho(E;U)$ when $U(0)\not=0$, but this formula  (eqn. (\ref{toldox})) is much more complicated.

Some caveats are inevitable here.   There are choices to be made in determining which quantum theory corresponds to a particular
classical function $W(\phi)$.   Generically, in passing from classical to quantum mechanics, a family of classical systems
corresponds to a family of quantum systems, but because of issues of operator ordering and renormalization, there is generically no natural one-to-one mapping from classical systems to
quantum systems.    To illustrate the difficulty in the present context, consider an example that actually will be important in this paper, 
namely a simple exponential
perturbation, $U(\phi)=2\veps e^{-\alpha\phi}$.  The perturbation of the action is $-\veps \int \d^2x \sqrt g e^{-\alpha\phi}$.   
Suppose that we  look at JT gravity in conformal
gauge with $g_{\mu\nu}=\delta_{\mu\nu}e^{2\sigma}$.   The action becomes $\int\d^2x\left(\phi \partial_\alpha\partial^\alpha\sigma
-e^{2\sigma}\phi\right)$, and the perturbation is $\int \d^2x e^{2\sigma-\alpha\phi}$.   With this action, at the very least
 there is a $\langle \phi\sigma\rangle$
two-point function, so the operator $e^{2\sigma-\alpha\phi}$ certainly needs renormalization and cannot just be treated classically.   
Different renormalization procedures will associate different quantum operators to the same classical expression 
 $\int \d^2x \sqrt g \,e^{-\alpha\phi}$.

When we go to second order and higher orders in $U$, matters become more complicated.   Perturbation theory involves multiple integrals such as $\int \d^2x_1 \sqrt{g(x_1)}\,e^{-\alpha\phi(x_1)}\int \d^2x_2\sqrt{g(x_2)} \,e^{-\alpha\phi(x_2)}$.   
In quantum field theory, more input is needed to define such a product.
Products such as  \be\label{wiff} \sqrt{g(x_1)}\,e^{-\alpha\phi(x_1)}\cdot \sqrt{g(x_2)} \,e^{-\alpha\phi(x_2)}\ee are only well-defined modulo possible
contact terms.  There is no way to fix the contact terms 
 until one specifies exactly
how one is trying to parametrize a family of quantum theories by a given family of classical field theories.   See  \cite{Kutasov}
for  a discussion of this in the context of the moduli space of two-dimensional conformal field theories.

In this article, we adopt a particular way to resolve all these questions, which makes possible relatively simple calculations.
This procedure will be based on a correspondence between an exponential perturbation $e^{-\alpha\phi}$ and a conical singularity
with deficit angle $\alpha$.    
Our detailed formulas depend on the procedure we adopt and are valid in the context of that procedure.  But a different procedure
would lead to different formulas.    An invariant statement
is that for any $W(\phi)$, the theory is dual to a hermitian matrix model.  The correspondence between functions $W(\phi)$
and eigenvalue densities $\rho(E)$ characterizing the dual matrix model is scheme-dependent.  

There actually is some further fine print.   We do not literally calculate for an arbitrary perturbation $U(\phi)$, 
but for a perturbation of the form
\be\label{wytop} U(\phi)=2\sum_{i=1}^r\veps_i \exp(-\alpha_i\phi),~~ \pi<\alpha_i<2\pi. \ee
For technical reasons, it is more simple to calculate if the exponents $\alpha_i$
 are in the indicated range.   The discrete sum in eqn. 
(\ref{wytop}) can also be replaced by (or supplemented  with) an integral.   However, one would expect  perturbation theory to be holomorphic in $U$ (after restricting the behavior of $U(\phi)$ for large $\phi$ in a way that is needed for perturbation
theory to make sense), so it is likely that our main results, such as eqn. (\ref{onojjo}),  hold for much more general $U$.

A more serious issue is that our results are valid to all orders of perturbation theory (at least for the class of functions
we consider), but there may be nonperturbative phenomena.   In fact, there must be some sort of nonperturbative phenomenon.    The function $\rho(E;U)$ in eqn. (\ref{onojjo}) and likewise its generalization in eqn. (\ref{toldox}) is positive for all $E>E_0$
if $U$ is sufficiently
small, but positivity can be lost if $U$ becomes large.    One possibility is that  nonperturbative corrections  to eqns. (\ref{onojjo}) and
(\ref{toldox})  ensure positivity for all $U$.   However, another interpretation seems more likely.
It is tempting to suspect that  these formulas are actually correct if
$U$ is sufficiently small, but break down at a  phase transition when (or possibly before) positivity is lost.   What makes
this plausible is that something similar happens classically.    In classical physics, the ground state  in a model of this type 
can jump discontinuously at a first-order phase transition when $U$ is varied \cite{EW}.
A quantum analog of this might be a  phase transition that prevents $\rho(E;U)$ from becoming negative.     
A failure of positivity might be avoided via a first-order phase transition if $\rho(E;U)$ changes discontinuously with $U$ in a suitable fashion,
or via a second-order phase transition if $\rho(E;U)$ changes continuously but not in a real-analytic fashion.    The classical picture
suggests first-order transitions.

JT gravity is an example of a ``single-cut'' matrix model -- a model in which the eigenvalue density in the large $N$ limit (or the large $\SS_0$ limit\footnote{$\SS_0$ is the ground state entropy.   After double-scaling, the large $N$ limit becomes the large $\SS_0$ limit.})
  is supported on a connected
interval, in this case $[0,\infty)$.   There are also ``multi-cut'' matrix models in which the eigenvalue density is supported on a union of
disjoint intervals,\footnote{To explain the terminology, in a hermitian matrix model, the
 trace of the resolvent $R(E)=\Tr\,\frac{1}{E-H}$ is holomorphic in $E$ with cuts on the real axis on the support of the
function $\rho(E)$.   The number of intervals on which $\rho(E)$ is supported therefore equals the number of cuts.}
 for example $[E_0,E_1]$ and $[E_2,\infty)$ with $E_0<E_1<E_2$.     It is tempting to suspect that JT gravity
is separated by a  phase transition from a family of two-cut matrix models.   Then by extension, further phase transitions might
lead to models with any number of cuts.   Classically, for suitable $U$, there are gaps in the possible energies of black hole solutions \cite{EW}.
It would seem that the matrix model counterpart of a model with a gap in the black hole spectrum
 should be a model with gaps in the support of $\rho(E;U)$; in other
words, it should be a multi-cut matrix model.

The contents of this article are as follows.  In sections \ref{sectwo}, \ref{background}, and \ref{schwarzian} we explain the strategy of the
calculation and some background.  The basic idea is that with the type of perturbation we consider, perturbative contributions
to the path integral can be evaluated in terms of Weil-Petersson volumes of moduli spaces of Riemann surfaces with conical 
singularities.  This generalizes the relationship between JT gravity and Weil-Petersson volumes that was exploited in \cite{SSS}.
 In section \ref{firstorder}, we compute the correction to the eigenvalue density $\rho(E;U)$ that
is of first order in $U$. In section \ref{higher}, we explain that the first order result is actually the whole answer if $U(0)=0$. The invariant
meaning of the condition $U(0)=0$ is that the perturbation does not cause a shift in the ground state energy.  In section
\ref{exrel}, we drop the assumption $U(0)=0$ and compute the terms in $\rho(E;U)$ that are
 of second and third order in $U$.   The hypothesis that the
model can be described by a matrix model passes a nontrivial test:    the second and third order terms in $U$ have the properties
needed to reproduce the expected threshold behavior $\rho(E)\sim (E-E_0)^{1/2}$ of a hermitian matrix model.     In section \ref{exth},
we show that if one assumes that the model can be described as a matrix model, then one can determine exact formulas for
the threshold energy $E_0(U)$ and the eigenvalue density $\rho(E;U)$.   Once $\rho(E;U)$, which is determined by the path
integral on a disc, is known, the path integral on a two-manifold of any other topology is uniquely determined, assuming that the model
is indeed described by a hermitian matrix model.   So by computing for other topologies, one can test the matrix model hypothesis.
     In section \ref{hco}, we carry out a number of such tests by computing perturbative corrections on a sphere with two or three holes
     and on a one-holed torus.  In all cases, the gravity results agree with expectations from a hermitian matrix model.

     A paper developing a similar analysis of deformed JT gravity as a matrix model and placing this in an interesting context
     of three-dimensional gravity has appeared recently \cite{MaxfieldTuriaci}.   The authors also draw on some prior computations
    \cite{MertensTuriaci,MertensTuriaci2}.
     Many  facts about surfaces with conical singularities that we use in this paper were  also used in \cite{CJM} in an application
     to JT gravity in de Sitter space.
     
     Another although rather  formal way to connect  the deformations of JT gravity that we consider here to
     hermitian matrix models would go as follows.   Two-dimensional gravity defined via matrix models is related to intersection
     theory on the moduli space of Riemann surfaces \cite{WittenOld}, which in this context has been called topological gravity.
     The basic relationship is that the expectation value of a product of matrix traces can be expressed in terms of
     the tautological intersection numbers that are defined in eqn. (\ref{jugo}) below (these have been called the correlation functions
     of topological gravity).   The Weil-Petersson volumes of moduli space can also be expressed in terms of the same intersection
     numbers.   The procedure for doing so was reviewed in \cite{DW}, and as explained in that paper (see sections 2.4 and 4.2)
     this gives one way to derive the spectral curve, originally found by Eynard and Orantin \cite{EO2}, that is related
     to Weil-Petersson volumes.   Eqn. (\ref{nugo}) below, which is a formula of Mirzakhani \cite{M} later extended by Tan, Wong, and Zhang
     to surfaces with conical singularities \cite{TWZ}, shows that the volumes of surfaces with conical singularities are also part of the
     same intersection theory.   So these volumes, too, will be associated to a spectral curve.   It should be possible to recover the 
     results of section \ref{exth} from that point of view.

\section{Exponential Perturbations and Conical Singularities}\label{sectwo}

We write 
\be\label{zowrite} W(\phi)=2\phi+U(\phi),\ee
where $U=0$ corresponds to JT gravity.
It turns out that it is convenient to begin by considering the perturbation $U=2{\varepsilon}\, e^{-\alpha \phi}$, with constants
$\alpha$, $\veps$.  To start with, we work to order $\veps$, and then we will  learn how to make exact statements.

First let us discuss the most naive way to compute effects of order $\veps$.   
Including the perturbation, the bulk action is
\be\label{Taction}I= I_\JT -\veps\int \d^2x\sqrt g e^{-\alpha\phi}
. \ee
As usual, the argument of the Euclidean path integral is $\exp(-I)$.  Expanded in perturbation theory in $\veps$, this is 
\begin{align}\label{waction}\exp(-I)=\exp(-I_\JT)\left(1+\veps\int\d^2x_1\sqrt g \,e^{-\alpha\phi(x_1)}+\frac{\veps^2}{2} \int\d^2x_1\sqrt g \,e^{-\alpha\phi(x_1)}\int\d^2x_2\sqrt g \,e^{-\alpha\phi(x_2)}
+\cdots\right). \end{align}
The most obvious starting point to
evaluate the effects of the perturbation in order $\veps$ is to just evaluate the integral $\int\d^2x \sqrt g e^{-\alpha\phi}$
in the background classical solution.

The metric of $\AdS_2$ can be written:
\be\label{mestads}\d s^2= \d \rho^2+\sinh^2\rho\, \d\psi^2,~~\psi\cong\psi+2\pi. \ee   However, one usually wants to study JT
gravity on a large portion of $\AdS_2$, not on all of it \cite{MSY}.   Usually one considers a disc $D$ embedded in $\AdS_2$ such
that the circumference of the boundary has a prescribed value $L$, and the dilaton field $\phi$ on the boundary has a large value $\phi_\b$.
Then one takes $L$ and $\phi_\b$ to infinity with fixed ratio.
The classical equations for the dilaton field in JT gravity are satisfied by
\be\label{noko}\phi=\sC  \cosh\rho,\ee
for a constant $\sC $.   Any other solution for which $\phi$ grows everywhere at infinity
is related to this one by a  ${\mathrm{PSL}}(2,\R)$ transformation, so we might
as well use this solution.   If we do, then the condition that $\phi=\phi_\b$ on $\partial D$ tells us that $D$ must
be the region $\rho\leq \rho_0$ where
\be\label{loko} \phi_\b=\sC  \cosh\rho_0. \ee
The boundary of the region $D$ has circumference 
\be\label{poko} L =2\pi \sinh\rho_0, \ee
and we see that taking $L$ and $\phi_\b$ to infinity with fixed ratio can be achieved by simply taking $\rho_0$ large with fixed $\sC $.
The renormalized circumference of the disc is defined as
\be\label{fif}\beta=\lim_{\rho_0\to\infty} \frac{L}{2\phi_\b}.\ee
The factor of $1/2$ in this formula is an arbitrary normalization, which has been chosen to agree with \cite{SSS}.\footnote{\label{normalztn} See
eqn. (99) in \cite{SSS} and the following discussion.   Note also that as remarked after eqn. (4) in that paper, the typical choice made there is
$\gamma=1/2$, which corresponds to the factor of $1/2$ in eqn. (\ref{fif}).}   
Choosing this factor
amounts to defining the units of length and energy in the dual quantum mechanical theory.  With the chosen normalization, we have $\beta=\pi/\sC $,
so
\be\label{boswell}\phi=\frac{\pi}{\beta}\cosh\rho. \ee

With the $\AdS_2$ metric (\ref{mestads}) and the dilaton field (\ref{boswell}), it is straightforward to evaluate the order $\veps$ term
in eqn. (\ref{waction}).   We have to evaluate $\lim_{\rho_0\to\infty}\int_0^{\rho_0}\d\rho\,\sinh\rho \int_0^{2\pi}\d\psi \,e^{-\pi \alpha\cosh\rho/\beta}$.
This limit only exists if $\alpha>0$, so we restrict ourselves to this case.
With this restriction, the limit is $\frac{2\beta}{\alpha}\exp(-\pi\alpha/\beta)$.    Therefore, in this approximation the partition function of the perturbed theory on the disc $D$ is 
\be\label{huccu}Z_D(U)= Z^\JT_{D}\left(1+\veps\cdot \frac{2\beta}{\alpha} \exp\left(-\frac{\pi \alpha}{\beta}\right)+\O(\veps^2)\right), \ee
where $Z_{D}^\JT$ (to be discussed later) is the disc partition function in JT gravity.   

Even in linear order in $\veps$, this is not an exact answer.   It is valid for small $\alpha$, but to the extent that $\alpha$ is not
small, we need to worry about quantum fluctuations 
around the background classical solution.   As discussed in the introduction, to calculate the effects of quantum fluctuations, one
first needs to specify a procedure for defining the theory -- a cutoff or renormalization procedure that enables one to define the
operator $\sqrt g e^{-\alpha\phi}$.
As also discussed in the introduction, even after the term linear in $\veps$ is defined, to define the higher order terms requires
 more input about the passage from
classical physics to quantum theory.   That is because even once the operator $\sqrt g \, e^{-\alpha\phi}$ is defined, it takes more input
to fix possible contact terms in integrated operator products such as 
$\int\d^2x_1\sqrt g \,e^{-\alpha\phi(x_1)}\int\d^2x_2\sqrt g \,e^{-\alpha\phi(x_2)}$.

We will follow a specific procedure for resolving all these questions.   This procedure
 will make everything well-defined and effectively calculable.   The procedure
amounts to a specific way to define all of the operators and operator products that will appear.   Unfortunately, it is not extremely clear
how to match this procedure with the classical formula (\ref{huccu}).   After some work (see eqn. (\ref{dofo})), we will find that the coefficient
 of the $\O(\veps)$ term in the approach that we will follow is not the classical result $\frac{2\beta}{\alpha} \exp\left(-\frac{\pi \alpha}{\beta}\right)$
 but is
\be\label{tofo}2\beta \,\exp\left(\frac{-\pi\alpha+\alpha^2/4}{\beta}    \right) .\ee
The shift in the exponent from $-\pi\alpha$ to $-\pi\alpha+\alpha^2/4$ can presumably be interpreted as  a normal-ordering effect
resulting from a connected correlator  $\langle \phi\phi\rangle$.   The disappearance of the prefactor $1/\alpha$ in the classical formula
is harder to interpret, beyond saying that it is a renormalization factor for the operator $\sqrt g \,e^{-\alpha\phi}$.

Now we will describe the procedure that we will follow to define these integrals.  Let us go back to the term of order $\veps$ in eqn. (\ref{waction}).
  Pulling the integral over $x_1$ outside,
the path integral that computes the $\O(\veps)$ term is
\be\label{ction}\frac{1}{\mathrm{vol}}\int \d^2 x_1 \sqrt {g(x_1)}  \int D\phi\,Dg\, \exp\left(\frac{1}{2}\int \d^2x \sqrt g\phi(R+2) -\alpha \phi(x_1)\right).\ee
Here $\mathrm{vol}$ is formally the volume of the diffeomorphism group.  
To avoid subtleties involving the Schwarzian mode on the boundary, suppose for a moment that we are on a compact Riemann surface $\Sigma$
without boundary.   The path integral of JT gravity is usually evaluated
by integrating first over $\phi$ (after a contour rotation $\phi\to \i\phi$).   This usually gives a delta function supported at $R+2=0$.   
The constraint $R+2=0$ fixes the Weyl factor in the metric of $\Sigma$, and the space of metrics that satisfy this constraint,
modulo diffeomorphisms, is the moduli space of Riemann surfaces.

But here we are integrating over the choice of a metric on a surface together with a point $x_1$ on the surface.  
Moreover, there is an extra term  $\alpha \phi(x_1)$  in the exponent of the path integral.    The constraint that comes from
integration over $\phi$ is now
\be\label{woddor} R(x)+2=2\alpha \delta^2(x-x_1) \ee
 (where $\delta^2(x-x_1)$ is a delta function supported at $x=x_1$ and normalized to $\int \d^2x \sqrt g \delta^2(x-x_1)=1$). 
 This equation implies that the metric has a conical singularity at $x=x_1$, with deficit angle $\alpha$.   
 In other words, in polar coordinates centered on $x=x_1$, the geometry can be modeled locally by 
\be\label{wofo}\d s^2=\d r^2+r^2\d\varphi^2,~~~ \varphi\cong\varphi+2\pi-\alpha. \ee   
The constraint (\ref{woddor}) still uniquely fixes the Weyl factor in the metric of $\Sigma$, provided that $\alpha<2\pi$.
(This condition is needed because there is no real classical geometry with deficit angle greater than $2\pi$.)   So the space of
metrics that satisfy this constraint, modulo diffeomorphisms, is the same as the moduli space of complex Riemann surfaces 
 with a marked point or puncture.  If $\Sigma$ is a Riemann surface of genus $g$, this moduli space is usually denoted as
$\M_{g,1}$.

In order $\veps^k$, the action has an extra term $\alpha\sum_{i=1}^k \phi(x_i)$, and 
 the integration over $\phi$ will give
  \be\label{oddor} R(x)+2=2\alpha \sum_{i=1}^k \delta^2(x-x_i),\ee
  leading to conical singularities with deficit angle $\alpha$ at each of the $k$ points $x_1,x_2,\cdots,x_k$.
The constraint still fixes the Weyl factor of the metric, and the space of metrics that satisfy this constraint modulo
diffeomorphisms is now
$\M_{g,k}$, the moduli space of Riemann surfaces of genus $g$ with $k$ punctures.   $\M_{g,k}$ is
defined, to be more precise, as the moduli space of Riemann surfaces of genus $g$ with $k$ {\it distinct} punctures.   So in
an approach that leads to a convergent integral over $\M_{g,k}$, the punctures will never collide and therefore the question of
contact terms in multiple integrals will be avoided.   Thus, such an approach gives a way to resolve, or avoid, all of the questions
about contact terms.    It gives us a specific way to map a family of classical theories to a family of quantum theories.

In all cases, the constraint says that $R+2=0$ away from conical singularities.    A Riemann surface with a metric that satisfies
$R+2=0$, away from possible conical singularities, is called a hyperbolic  Riemann surface.    So in the present context, it is natural
to think of $\M_{g,k}$ as parametrizing a family of hyperbolic Riemann surfaces, in this case with a conical singularity of deficit angle $\alpha$
at each of the $k$  punctures or marked points.

In evaluating the path integral, each conical singularity comes with a factor $\veps$.   
  A contribution with $k$ conical singularities has an additional factor $1/k!$.   The meaning of this factor is that the conical
  singularities should be treated as indistinguishable; independent integration over positions $x_1,x_2,\cdots,x_k$  or in other
  words over the moduli space $\M_{g,k}$ (which is usually defined in terms of distinguishable, labeled punctures) gives an overcounting
  by a factor of $k!$.   
  
  Let $\M_{g,\vec\alpha}$ be
the moduli space of hyperbolic Riemann surfaces $\Sigma$ of genus $g$ with $k$ conical singularities with possibly different
deficit angles $\vec\alpha=(\alpha_1,\alpha_2,
\cdots,\alpha_k)$.  
    A Weil-Petersson symplectic form can be defined in the usual way on the moduli space $\M_{g,\vec\alpha}$, and the usual arguments concerning the path integral of JT gravity \cite{SSS,SW}
  show that the path integral on such a $\Sigma$  computes the volume of  $\M_{g,\vec\alpha}$.  
   
  We are actually mainly interested in the case that $\Sigma$ has asymptotically $\AdS$ boundaries.
In this case, as usual, the relation between the JT integral and the Weil-Petersson volume
 must be modified to incorporate a Schwarzian mode along
  each asymptotically AdS boundary. The reasoning that leads to this \cite{MSY,SSS} 
  is not affected by the possible presence of conical singularities in the interior of $\Sigma$.

In the classical computation  that we started with, the natural constraint on $\alpha$ was $\alpha>0$.   In the computation
that we will perform, based on the relation to Weil-Petersson volumes of Riemann surfaces with conical singularities,
the situation is slightly different.   
We have already remarked that  eqn. (\ref{oddor})  only has real solutions if $\alpha<2\pi$.
But  for a more subtle reason that will be explained momentarily, in our analysis we will also have to assume that $\alpha>\pi$.   Combining
these constraints, our calculations will be directly applicable only for $2\pi>\alpha>\pi$.    
 
   \begin{figure}
 \begin{center}
   \includegraphics[width=2.7in]{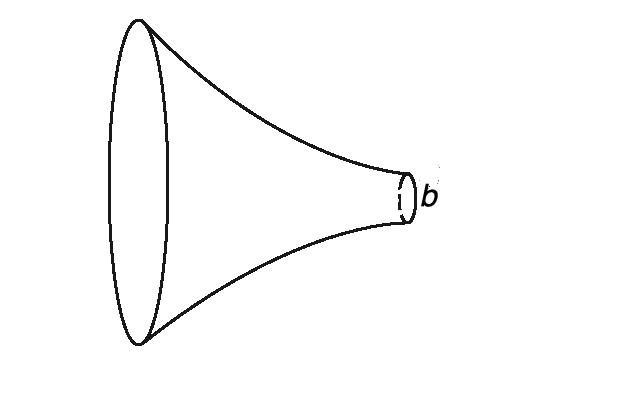}
 \end{center}
\caption{\small  By a ``trumpet,'' we mean a hyperbolic two-manifold that is topologically an annulus.   One ``inner'' boundary is
a geodesic, here of length $b$.   The other ``outer'' boundary is exceedingly large; it represents an approximation to the asymptotic
boundary of $\AdS_2$.
\label{Trumpet}}
\end{figure}

   \begin{figure}
 \begin{center}
   \includegraphics[width=2.7in]{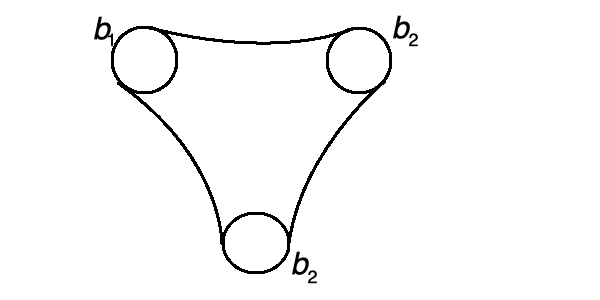}
 \end{center}
\caption{\small A three-holed sphere with a hyperbolic metric and with geodesic boundaries of lengths $b_1,b_2,b_3$.   By gluing
together three-holed spheres and/or trumpets along their geodesic boundaries, one can make more complicated hyperbolic two-manifolds.
A simple example involving gluing of a trumpet to a three-holed sphere is in fig. \ref{FigOne}(a).
\label{ThreeHoled}}
\end{figure}
 
  \begin{figure}
 \begin{center}
   \includegraphics[width=3.4in]{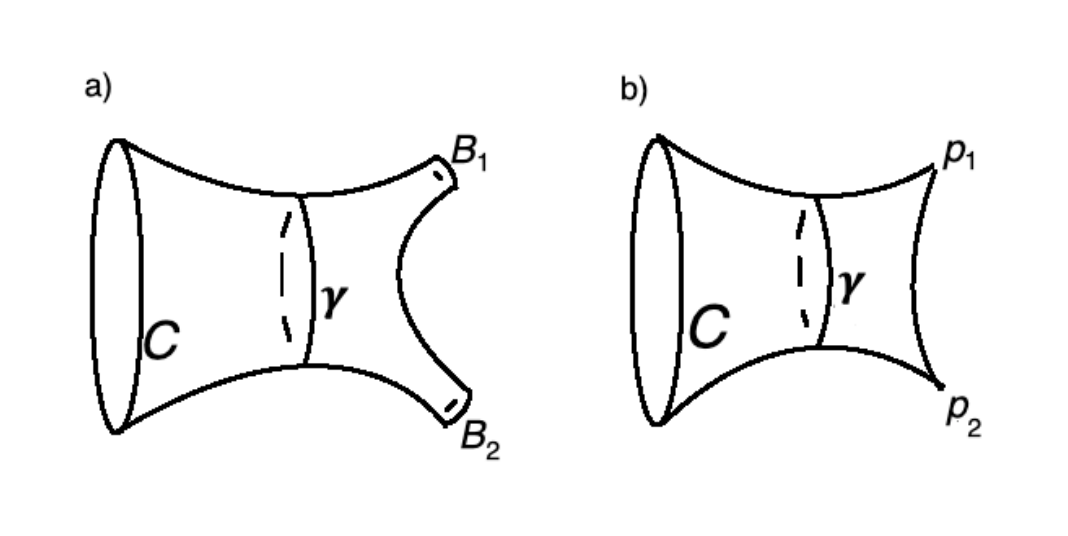}
 \end{center}
\caption{\small (a)   A Riemann surface $\Sigma$
 of genus 0 with an asymptotically AdS boundary $C$, and two geodesic boundaries $B_1$, $B_2$.
We consider loops homologous to $C$.   Minimizing the length of such a loop in its homology class, we arrive at the closed
geodesic $\gamma$. Once we find $\gamma$, $\Sigma$ can be reconstructed by gluing a trumpet and a three-holed sphere along
$\gamma$.   (b)   The geodesic boundaries of $\Sigma$ have been replaced by conical singularities $p_1$, $p_2$, with deficit angles
$\alpha_1$, $\alpha_2$.   Again we consider loops homologous to $C$.   As long as $\alpha_1,\alpha_2>\pi$, minimizing the length of
such a loop leads to a picture rather similar to that in (a), again with a closed geodesic $\gamma$.    For the reason for the constraint
$\alpha_1,\alpha_2>\pi$, see fig. \ref{FigTwo}.   $\Sigma$ can be reconstructed by gluing two simple-building blocks along $\gamma$.
One building block is a trumpet, and the other is similar to a three-holed sphere, but with two of the holes replaced by conical singularities.
\label{FigOne}}
\end{figure}

The constraint $\alpha>\pi$ arises is our calculations in the following way.
We will analyze
the moduli spaces assuming that conical singularities behave similarly to geodesic boundaries.   One of the ingredients in \cite{SSS} as
well as subsequent work \cite{SW} was that a hyperbolic Riemann surface can be decomposed into  more simple building blocks by cutting
on suitably chosen geodesics.    The important irreducible building blocks (for oriented surfaces) are a ``trumpet,'' which is
an annulus with an asymptotically AdS boundary  and a geodesic boundary (fig. \ref{Trumpet}), and a three-holed sphere with geodesic boundaries (fig. \ref{ThreeHoled}).     Riemann surfaces that do not have any asymptotically AdS boundaries are constructed
by gluing together three-holed spheres, without need for the trumpet.   The trumpet gives a way to incorporate asymptotically AdS boundaries.

An example of building a more complicated surface from these building blocks is sketched in fig. \ref{FigOne}(a).   Here, we consider
a surface $\Sigma$ of genus 0 that has an asymptotically AdS boundary $C$ and two geodesic boundaries $B_1$, $B_2$.  Consider
a loop in $\Sigma$ that is homologous to $C$.  By minimizing its length in its homology class, one can find a closed geodesic
$\gamma$, as sketched in the figure.   Then by cutting $\Sigma$ along $\gamma$, we decompose it into two elementary
building blocks, namely a trumpet and a three-holed sphere.   This is a typical example of decomposing a hyperbolic Riemann
surface into basic building blocks by cutting on  closed geodesics.  In this particular case, only one cut and two building blocks are needed.

 \begin{figure}
 \begin{center}
   \includegraphics[width=3.4in]{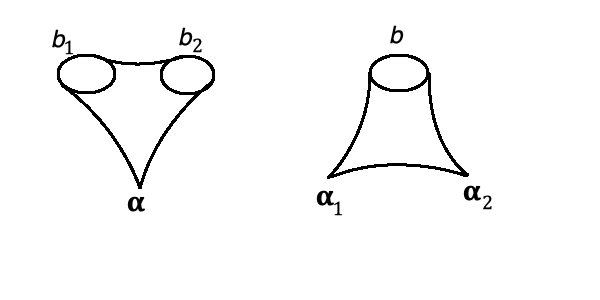}
 \end{center}
\caption{\small In the presence of conical singularities, hyperbolic Riemann surfaces can still be built by gluing together
elementary building blocks along their geodesic boundaries, but one has to include two additional elementary building blocks.
   The new building blocks are a sphere with two geodesic
boundaries and one conical singularity or with one geodesic boundary and two conical singularities, as pictured here. The geodesic
boundaries are labeled by their circumferences and the conical singularities are labeled by their deficit angles. \label{Examples}}
\end{figure}

 \begin{figure}
 \begin{center}
   \includegraphics[width=3.4in]{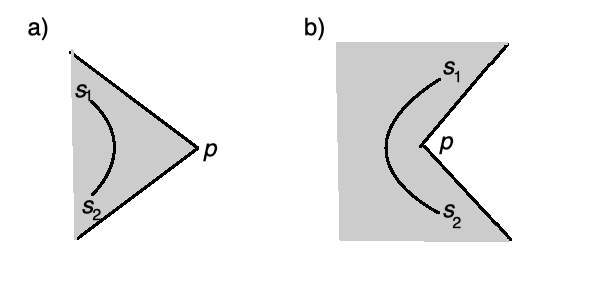}
 \end{center}
\caption{\small  A Riemann surface $\Sigma$ with a conical singularity at a point $p$ has been, locally, ``unwrapped'' to a wedge-shaped
region of the plane. (The black rays emanating from $p$ should be identified to build $\Sigma$.)   The curvature of $\Sigma$ does not affect the issue that will be discussed here, because it is unimportant
in a small neighborhood of $p$; it is neglected in this figure.   (a)   The deficit angle at $p$ is greater than $\pi$.   Consider
a  curve passing between specified endpoints $s_1, s_2$ on a given side of $p$.
  In minimizing the length of a path from $s_1$ to $s_2$, the path  is ``repelled'' from p.   There is always
a geodesic between $s_1$ and $s_2$ on any pre-chosen side of $p$.  (As an exercise, the reader can try to describe the
geodesic from $s_1$ to $s_2$ that goes around $p$ on the other side; one has to use the fact that the two black rays from $p$
are identified.)  (b)  The deficit angle at $p$ is less than $\pi$.   A geodesic with specified endpoints does not necessarily
exist on a given side of $p$.   For example, in the figure, to shorten the indicated path from $s_1$ to $s_2$ to a geodesic,
one would have to pull it across the point $p$.  \label{FigTwo}}
\end{figure}

We would like to treat hyperbolic Riemann surfaces with conical singularities  similarly.   In this case, in addition to 
the trumpet and the three-holed sphere, we  have to allow two more basic building blocks, which are obtained
by replacing  one or two of the geodesic boundaries of a three-holed sphere  with conical singularities.  See fig. 
\ref{Examples}.\footnote{\label{exceptional} We do not add a sphere with three 
conical singularities as an additional building block because, as it has
no boundary, it cannot be used in any further gluing.   But because the sphere with three conical singularities
cannot be made from the other building blocks, it is an exceptional case in many statements.  For example,
that is why, when we discuss a specific example in section \ref{background}, we will consider a sphere with four conical
singularities, to which the general theory applies.}
    As a typical
example of decomposing a surface with conical singularities into these more general elementary building blocks, let us consider a surface
$\Sigma$ that is a disc with an asymptotically AdS boundary and two conical singularities (fig. \ref{FigOne}(b)).     In other
words, we replace the geodesic boundaries $B_1,B_2$ in fig. \ref{FigOne}(a) with conical singularities $p_1$, $p_2$ in 
fig \ref{FigOne}(b).   Now we try to again decompose $\Sigma$ into elementary building blocks by minimizing the length of a loop
that is homologous to the outer boundary
 $C$. More specifically, we want to minimize the length of this loop without letting it jump across or pass through
any of the conical singularities.
If the minimizer exists, as sketched in fig. \ref{FigOne}(b),  we get the
desired decomposition: the building blocks are now  a trumpet and a sphere with one hole with geodesic boundary (namely $\gamma$)
and two conical singularities   ($p_1$ and $p_2$).    We will use such decompositions many times in our computations.

However, in general this process will only work if, when we try to minimize the length of a loop, that loop is ``repelled'' from any conical 
singularities.   If in the minimizing process, a curve gets pulled across a conical singularity,  we will not get the desired decomposition.
 As an example to show that a nontrivial condition is needed,
consider a disc with two conical singularities, as in  fig. \ref{FigOne}(b), but now assume that
the two deficit angles are very small.   In this case, the geometry of the disc is almost the same
as that of a disc in $\AdS_2$.   In $\AdS_2$, there are no closed geodesics, and adding conical singularities with very small deficit angle does
not bring closed geodesics into existence.  If we start with a loop homologous to $C$ and try to minimize its length, it will
just jump across the conical singularities and shrink to a point.
 So the deficit angles must be sufficiently large to generate the decomposition sketched
in fig. \ref{FigOne}(b).

In fact, the necessary condition to ensure that the desired minimizers always exist  
 is that the deficit angles
should be all greater than $\pi$.    The reason is illustrated in fig. \ref{FigTwo}.
 Consider two points $s_1,s_2\in\Sigma$.  Suppose that we are given a simple curve\footnote{A simple curve is a curve that
 does not intersect itself.  If a path $\gamma$ from $s_1$ to $s_2$ intersects itself at a point $p$, then by dropping from $\gamma$ a closed loop, we get a shorter path from $s_1$ to $s_2$.   So in looking for length-minimizing curves, it is reasonable to consider simple curves.}
 between $s_1$ and $s_2$  in the smooth part of $\Sigma$,
 and we want to shorten this curve to a geodesic.
Consider first fig. \ref{FigTwo}(a).   Here we illustrate a conical singularity with a deficit angle greater than $\pi$.
When one tries to minimize the length of a curve between given points $s_1$, $s_2$, it is ``repelled'' from the singularity,
in the sense that any curve from $s_1$ to $s_2$ can be shortened to turn it into a geodesic without pulling it
across the conical singularity at $p$.  
One can specify at will which way a simple curve from $s_1$ to $s_2$ should go around $p$, 
and a minimizing geodesic with that property will exist.   

 That is not true for a conical singularity with deficit angle less than $\pi$ (fig. \ref{FigTwo}(b)).
To shorten the indicated curve from $s_1$ to $s_2$, one would have to pull this curve across $p$, and one
would end up with a geodesic that goes around $p$ on the other side.

We have here described the case of curves with
specified endpoints,  but 
the case we actually need of closed loops is similar.   The reason for this is that whether a curve is attracted to or repelled from a conical
singularity $p$ when we try to minimize its length depends only on local considerations near $p$, not on global properties of the curve.

In short, the techniques that we will use in this article to analyze the perturbation  $2\veps\exp(-\alpha\phi)$ will only be
valid in the range $\pi<\alpha<2\pi$.    However, the results that we will get will be analytic in $\alpha$, suggesting that they are valid
outside the range that we start in (perhaps even for negative\footnote{What happens for negative $\alpha$ is that the high energy
behavior of the eigenvalue distribution of the dual matrix model no longer coincides with that of JT gravity.  But a matrix model
with that distribution does exist, as long as the sign of the coupling $\veps$ is chosen so that the distribution is positive.}  $\alpha$).   
More generally, by the same methods we will
be able to analyze a linear combination of such exponentials,
\be\label{weldo}U(\phi)=2\sum_i\veps_i \exp(-\alpha_i\phi),~~ \pi<\alpha_i<2\pi. \ee
The result that we will get for the eigenvalue density of a dual matrix model can be expressed in a way that makes sense
for a much larger class of functions, so it probably is valid in more generality than the derivation.  

\section{Some Background On Conical Singularities And Volumes}\label{background}

 Volumes of
moduli spaces of Riemann surfaces with conical singularities, for the case of deficit angles greater than $\pi$, have been
analyzed in \cite{TWZ}.    Understanding how this was done is important background for the problem that we will be studying.

First of all,  let us compare conical singularities to holes with geodesic boundary.   
We  recall that a hyperbolic Riemann surface $\Sigma$ has
a natural flat $\mathrm{PSL}(2,\R)$ connection.   It is convenient  to lift the holonomies to $\SL(2,R)$ and
describe them as $2\times 2$ matrices of
determinant 1,  but this involves an arbitrary choice of sign.   The holonomy around a geodesic of length $b$ is conjugate (up to sign)  to
\be\label{pollo} U_b=\exp\begin{pmatrix}b/2& 0\cr 0 & -b/2\end{pmatrix}. \ee
What is the holonomy around a conical singularity with deficit angle $\alpha$?   In going around the conical singularity, a tangent vector
is rotated by an angle $\alpha$.   But $\SL(2,\R)$ acts on the spin bundle, and the rotation of a spinor is by an angle $\alpha/2$.
So, taking $\begin{pmatrix} 0 & 1\cr -1 &0\end{pmatrix}$ as the rotation generator, the monodromy of the flat connection around a conical
singularity with deficit angle $\alpha$ is
\be\label{ollo} V_\alpha =\exp\begin{pmatrix}0& \alpha/2\cr -\alpha/2 & 0\end{pmatrix}. \ee
We see that $U_b$ and $V_\alpha$ are conjugate in $\SL(2,\C)$ (up to sign) if we identify $b$ with $\i(2\pi-\alpha)$. In other words, at least from the point of view of the monodromy, a conical singularity can be viewed as the analytic
continuation of a hole with geodesic boundary.

From what we have said so far, it is not obvious why $b$ should be identified with $\pm \i(2\pi-\alpha)$ and not just
$\pm \i\alpha$.   Probably the most direct way to understand this is the following.   In the limit $b\to 0$, a geodesic boundary of length
$b$ reduces to a puncture, which in the context of hyperbolic geometry means a cusp.    Likewise, in the limit that $\alpha$ approaches
$2\pi$, a conical singularity with a deficit angle of $2\pi$ approaches a cusp.   So $b=0$ corresponds to $\alpha=2\pi$, not $\alpha=0$.
Note that in the limit $\alpha\to 0$, a conical singularity becomes a smooth point, not a cusp.   We will see in many other ways the
importance of including the $2\pi$ in the relation between $b$ and $\alpha$.

We will now make a small aside and explain something that is important in a theory with fermions  but not
 in a purely bosonic theory such as JT gravity.
A comparison between conical singularities and geodesic boundaries  that does include the signs of the holonomies
 can be made if we pick a spin structure, in which case the holonomies become well-defined
in $\SL(2,\R)$.   The generalization of eqn. (\ref{pollo}) is then $ U_b=\eta\exp\begin{pmatrix}b/2& 0\cr 0 & -b/2\end{pmatrix}$, where $\eta=-1$
for Neveu-Schwarz (NS) spin structure on the geodesic
circle, and $+1$ for Ramond (R) spin structure.   We then see that $V_\alpha$ is conjugate
in $\SL(2,\C)$ to
the analytic continuation of $U_b$
if the spin structure on the geodesic is of NS type. Here it is important that $b=\pm\i(2\pi-\alpha)$, not $\pm\i\alpha$.
If  the spin structure of the geodesic is of R type,
we need an extra minus sign in $V_\alpha$, which means that the conical singularity should be a Ramond singularity,
producing an extra minus sign for fermions.

This relationship between
conical singularities and geodesic boundaries has been important in understanding Weil-Petersson volumes in the presence of conical
singularities.
   Mirzakhani \cite{M} developed a technique to compute volumes of moduli spaces of hyperbolic 
Riemann surfaces, in general with geodesic boundaries.    Thus in general she computed the volume of the moduli space $\M_{g,\vec b}$
of hyperbolic Riemann surfaces of genus $g$ with geodesic boundaries of lengths $\vec b=(b_1,b_2,\cdots,b_n)$.
One important ingredient in her work was the fact that a hyperbolic
Riemann surface, possibly with geodesic boundaries,
 can be constructed (in many ways)  by gluing together three-holed spheres along geodesic boundaries.  
In \cite{TWZ}, it was shown that all the facts used by Mirzakhani have analogs for surfaces that have conical singularities  in addition to (or instead
of)  geodesic boundaries, as long as all the conical singularities have deficit angles in the range\footnote{In their terminology, the cone angle should be less than $\pi$; what they call the cone
angle is the opening angle $2\pi-\alpha$ at a conical singularity, where $\alpha$ is the deficit angle.}  $\pi<\alpha<2\pi$.
The restriction to $\alpha>\pi$ was needed
because of the phenomenon that we explained in discussing fig. \ref{FigTwo}: the decomposition of a surface with conical
singularities in elementary building blocks works the same way as for a surface with geodesic boundaries  if and only if the
deficit angles satisfy $\pi<\alpha<2\pi$.     The elementary building blocks that one has to allow are  familiar from section \ref{sectwo}:\footnote{
 If
one wants asymptotically AdS boundaries -- as we will -- one adds the trumpet as another building block.}
a three-holed sphere; a sphere with two holes and one conical singularity; and a sphere with one hole and two conical singularities.\footnote{  A sphere with precisely three conical singularities is an exceptional case because it cannot be constructed from the usual building blocks, as noted in footnote \ref{exceptional}.   As a result,
in this particular example, the condition $\alpha_i>\pi$ actually does not suffice for establishing a relationship between Weil-Petersson volumes
for surfaces with conical singularities and Weil-Petersson volumes for surfaces with holes.  A necessary and sufficient condition is that
the moduli space $\M_{0,\vec\alpha}$ should be non-empty.   It then consists of a single point and has the same volume -- namely 1 -- as
the moduli space for a three-holed sphere, which also consists of a single point.}
More specifically, it was shown in \cite{TWZ} that as long as deficit angles are in the favored range,
volumes of moduli spaces
of hyperbolic surfaces with conical singularities can be obtained from the volumes of hyperbolic surfaces with geodesic boundaries
by the analytic continuation $b\to \i (2\pi-\alpha)$.  This consequence of their results is stated in the Additional Remark at the
end of the introduction of their paper.  See also \cite{ND} for some further developments.

The formulas obtained in \cite{TWZ} -- like the formulas that we will obtain in the rest of the present paper -- are analytic in $\alpha$ and may
retain some validity beyond the region $\pi<\alpha<2\pi$.    However, some restriction on the $\alpha$'s is really needed.
To see this, let us consider the example of a four-holed sphere with geodesic boundaries of lengths $b_1,b_2,b_3,b_4$.
The volume of the moduli space was computed by Mirzakhani and is 
\be\label{vf} V_{0,4}= 2\pi^2+\frac{1}{2}\sum_{i=1}^4 b_i^2. \ee By the result of 
\cite{TWZ}, if we make a replacement $b_i\to \i (2\pi -\alpha_i)$ for some (or all) of the $b_i$, the same formula gives the volume
of the moduli space of hyperbolic metrics on a sphere with $p$ holes with geodesic boundary and $q$ conical singularities, with $p+q=4$,
as long as the deficit angles are in the favored range $\pi<\alpha_i<2\pi$.

However, the formula is not valid in general outside of this range.   For example,   
for $p=0$, $q=4$, the Gauss-Bonnet theorem gives a formula
\be\label{mk}\int_{S^2} R = 8\pi -2\sum_{i=1}^4\alpha_i ,\ee
where the integral is taken on the smooth part of $S^2$.   This formula
cannot be satisfied by a hyperbolic metric (which has $R<0$ away from the conical singularities) 
if $\sum_{i=1}^4\alpha_i\leq 4\pi$.   So the moduli space of hyperbolic metrics
on a sphere with the specified conical singularities is empty unless $\sum_{i=1}^4\alpha_i>4\pi$.  The condition $\alpha_i>\pi$ that is needed
for the theorem of \cite{TWZ} to hold ensures that $\sum_{i=1}^4\alpha_i>4\pi$.   The moduli space is then nonempty and according to the
result of \cite{TWZ}, its volume is
$2\pi^2-\frac{1}{2}\sum_{i=1}^4(2\pi-\alpha_i)^2$.

Mirzakhani also found a formula expressing the Weil-Petersson volume $V_{g,\vec b}$ of 
 $\M_{g,\vec b}$ in terms of an integral over $\M_{g,n}$, the moduli space of
Riemann surfaces with $n$ punctures.  
Topologically, $\M_{g,\vec b}$ and $\M_{g,n}$ are equivalent, as explained for example in \cite{DW}.   But
the corresponding Weil-Petersson forms $\omega_{g,\vec b}$ and $\omega_{g,n}$ are not cohomologous.   The relation between
them involves the following.
If $p$ is a point in a Riemann surface $\Sigma$, then the cotangent bundle to $\Sigma$ at $p$ is a one-dimensional complex vector
space.    As the moduli of $\Sigma$ vary, this vector space varies as the fiber of a complex line bundle $\L$ over the modulli space of Riemann
surfaces.   In the case of $\M_{g,n}$, there are $n$ punctures $p_1,\cdots,p_n$, and therefore this construction gives $n$ complex
line bundles $\L_1,\L_2,\cdots,\L_n$. Each of these line bundles has a first Chern class; we define $\psi_i=c_1(\L_i)$.
Mirzakhani found the relationship between $\omega_{g,\vec b}$ and $\omega_{g,n}$:
\be\label{oppo}\omega_{g,\vec b}=\omega_{g,n}+\frac{1}{2} \sum_i b_i^2\psi_i. \ee
This relation between symplectic forms immediately leads to a statement about volumes:\footnote{Here $\bar\M_{g,n}$ is the Deligne-Mumford compactification of $\M_{g,n}$. Actually, in hyperbolic
  geometry, the cohomology classes have natural representatives that vanish at infinity.   If one uses those representatives, one can just
  integrate over $\M_{g,n}$ rather than its compactification.   Note that even after  the compactification to $\bar\M_{g,n}$, punctures
  never collide so questions about contact terms do not arise.} 
  \be\label{nk} V_{g,\vec b}=\int_{\bar\M_{g,n}} \exp\left(\omega_{g,n}+\frac{1}{2}\sum_i b_i^2\psi_i \right). \ee
 Since the $\psi_i$ are two-dimensional classes, the
  integral on the right hand side is manifestly a polynomial in the variables $b_i^2$, of degree $3g-3+n$ (the complex dimension
 of $\M_{g,n}$).   The bottom term in this polynomial, obtained by setting all $b_i$ to zero, is the Weil-Petersson volume of $\M_{g,n}$, and
 the top degree terms, which are what we get if we drop $\omega_{g,n}$ from the exponent on the right hand side of eqn. (\ref{nk}),
 are the correlation functions of two-dimensional topological gravity, in the sense of \cite{WittenOld}.  These are defined
 as
 \be\label{jugo}\langle \tau_{d_1}\tau_{d_2}\cdots\tau_{d_n}\rangle =\int_{\bar\M_{g,n}} \psi_1^{d_1}\psi_2^{d_2}\cdots \psi_n^{d_n}. \ee

It was shown in  \cite{TWZ} that the same formulas hold under the substitution $b\to \i(2\pi-\alpha)$ if we replace some
 or all of the geodesic boundaries with conical singularities in the preferred range ($\pi<\alpha<2\pi$).     Thus, if $\M_{g,\vec b,\vec\alpha}$
 is the moduli space of hyperbolic Riemann surfaces of genus $g$ with geodesic boundaries of lengths $\vec b=(b_1,b_2,\cdots,b_m)$ and conical
 singularities with deficit angles $\vec\alpha=(\alpha_1,\alpha_2,\cdots,\alpha_n)$, then the volume $V_{g,\vec b,\vec\alpha}$ is
 \be\label{nugo}V_{g,\vec b,\vec \alpha}=\int_{\bar\M_{g,m+n}} \exp\left(\omega_{g,m+n}+\frac{1}{2}\sum_{i=1}^m b_i^2\psi_i
 -\frac{1}{2}\sum_{j=1}^n(2\pi-\alpha_j)^2\t\psi_j \right). \ee
 We have denoted the $\psi$ classes associated to the holes as $\psi_i, \,i=1,\cdots,m$, and those associated to the conical
 singularities as $\t \psi_j$, $j=1,\cdots,n$.
 If $\vec b=0$ and all $\alpha_j$ are equal to $2\pi$, then $\M_{g,\vec b,\vec\alpha}$ reduces to $\M_{g,n+m}$ and 
 $V_{g,\vec b,\vec\alpha}$ reduces to the volume of $\M_{g,n+m}$, which is $\int_{\bar\M_{g+n}}e^{\omega_{g,n+m}}$.

\section{Evaluating The Schwarzian Path Integral}\label{schwarzian}

Before proceeding to analyze how the perturbation $U=2{\veps}\exp(-\alpha\phi)$ affects the eigenvalue
distribution of the matrix model, we need one more step.

An important ingredient in \cite{SSS} is the evaluation of the Schwarzian path integral in two special cases: a disc with nearly $\AdS_2$
boundary, and a trumpet, which is the same as an annulus with one nearly $\AdS_2$ boundary and one geodesic boundary.
To study JT gravity with conical singularities, a third example is very important.   This is a disc with nearly $\AdS_2$ boundary
and a single conical singularity.   

The Schwarzian action arises in this context as the action of JT gravity, including the Gibbons-Hawking-York surface term,
in a cutoff version of a spacetime that is asymptotic to $\AdS_2$ \cite{MSY}.    The resulting path integral is the exponential of the
classical action times a one-loop determinant.    There are no higher corrections, as the Schwarzian path integral is one-loop exact
\cite{SWfirst}.

Let us first first review the calculation of the Schwarzian path integral for a disc $D$, which in this context is simply $\AdS_2$ itself
with a cutoff at large distances.   The bulk action $I_\JT$ of eqn. (\ref{JTaction}) vanishes on-shell, since the classical
equation of motion for $\phi$ gives $R+2=0$.     However, this bulk action requires a Gibbons-Hawking-York boundary term,
which is 
\be\label{GHY}I_\GHY=-\int_{\partial\Sigma}\d x \sqrt h \phi (K-1). \ee
where $h$ is the induced metric of the boundary and $K$ is its extrinsic curvature.   Using here $K-1$ rather than $K$ is not important
for getting a well-defined variational problem and quantum theory.   However, it is important for getting a theory that has a nearly
$\AdS_2$ limit.   The reason is simply that the extrinsic curvature of a  disc in $\AdS$ space approaches 1 in the limit of a large disc.

As was explained in section \ref{sectwo},  the metric of $\AdS_2$ is described by
\be\label{metads}\d s^2= \d \rho^2+\sinh^2\rho\, \d\psi^2,~~\psi\cong\psi+2\pi, \ee
and the usual boundary conditions for a large disc  embedded in $\AdS_2$ are satisfied by the disc $D$ defined by
$\rho\leq\rho_0$, with the dilaton
\be\label{thed}\phi=\frac{\pi}{\beta}\cosh \rho,\ee
where $\beta$ is the renormalized circumference of $D$.   
To evaluate the GHY surface term in this situation, we note that
the boundary of $D$ has circumference $ L =2\pi \sinh\rho_0.$
For $\phi|_{\partial D}$ equal to the constant $\phi_\b$, one term in the GHY action is
\be\label{joho}\int_{\partial D} \d x\sqrt h \phi = \phi_\b L= \frac{2\pi^2}{\beta}\cosh\rho_0\sinh\rho_0 . \ee
The other term is proportional to the extrinsic curvature.   One way to evaluate that term \cite{Yang} is to use the Gauss-Bonnet theorem
\be\label{oho}\int_{\partial D}\d x \sqrt h K=2\pi\chi-\frac{1}{2}\int_D\d^2x\sqrt g R,\ee
where $\chi$ is the Euler characteristic.   Since the disc has $\chi=1$ and $\AdS_2$ has $R=-2$, the right hand side of (\ref{oho})
is $2\pi +A$, where $A=2\pi (\cosh \rho_0-1)$ is the area of $D$.   So
\be\label{zoho}\int_{\partial D}\d x\sqrt h \phi_\b K = \frac{2\pi^2}{\beta} \cosh^2\rho_0.\ee 
Finally we see that the on-shell boundary action of a disc is $ -\frac{2\pi^2}{\beta}(\cosh^2\rho_0 -\cosh\rho_0\sinh\rho_0) =-\frac{\pi^2}{\beta}(1+e^{-2\rho_0})$,
which in the limit $\rho_0\to\infty$ becomes 
\be\label{nGHY} I_D=-\frac{\pi^2}{\beta}.  \ee

In the present paper, we will need the corresponding formula for what we will call $D(\alpha)$, a disc with a single conical singularity
with deficit angle $\alpha$.   The metric of $D(\alpha)$ can be described by the same formula (\ref{metads}), but now the range
of $\psi$ is $0\leq\psi\leq 2\pi-\alpha$ rather than $0\leq \psi\leq 2\pi$.    As before, we can satisfy the equation of motion
for $\phi$ and the boundary conditions by taking $\phi=\sC \cosh\rho$ and
 defining $D(\alpha)$ as the region $\rho\leq \rho_0$.    Because $\psi$ is now integrated
over a smaller range, we get now $L=(2\pi-\alpha)\sinh \rho_0$ and $\int_{\partial D(\alpha)} \d x \sqrt h K=(2\pi-\alpha)\cosh\rho_0$.
So (\ref{nGHY}) is replaced by $I_{D(\alpha)}=-(2\pi-\alpha)\sC /2$.   The formula for the renormalized circumference
$\beta$ becomes $\beta=\lim_{\rho_0\to\infty} L/2\phi_\b=(2\pi-\alpha)/2\sC $,
so in terms of $\beta$, we get
\be\label{zif} I_{D(\alpha)} =- \frac{(2\pi-\alpha)^2}{4\beta}. \ee

We will also need the corresponding formula for the trumpet, so let us recall how it is derived.   For the metric of a trumpet $T(b)$, we can take
\be\label{trumbo}\d s^2=\d \rho^2+\cosh^2\rho \,\d\psi^2,~~~ \psi\cong\psi+b.\ee
Here $\rho$ ranges over $0\leq \rho\leq\rho_0$, with a cutoff at very large $\rho_0$.  
The boundary at $\rho=0$ is a geodesic of circumference $b$.   The equation of motion
for the dilaton can be solved by $\phi = \sC \sinh \rho$.     Imposing that $\phi$ restricted to the boundary should be a constant $\phi_\b$,
we get $\phi_\b=\sC \sinh\rho_0$.    Imposing that the circumference should be $L$, we get $L=b\cosh\rho_0$.   Evaluating
$\int_{\partial T}\d x\sqrt h K$ as $2\pi\chi+A$, where $A$ is the area of $T$ and now $\chi=0$, we get $\int_{\partial T}\d x\sqrt h K=
b\sinh\rho_0$.     Hence the action of the trumpet is $I_T=\phi_\b(L-A)=\sC \sinh\rho_0 (b\cosh\rho_0-b\sinh\rho_0)=\frac{b\sC }{2}(1-e^{-2\rho_0})$.
Taking $\rho_0\to\infty$ and defining $\beta =\lim_{\rho_0\to\infty} L/2\phi_\b=b/2\sC $, we get
\be\label{traction}I_T=\frac{b^2}{4\beta}, \ee
Comparing eqns. (\ref{zif}) and (\ref{traction}), we see that the classical actions for $T(b)$ and $D(\alpha)$ are related by
$b\to \i (2\pi-\alpha)$, though the derivation did not make it obvious that this would be so. 

  The Schwarzian path integral is the exponential of the classical action times
a one-loop determinant. A localization argument shows that there are no higher order corrections \cite{SWfirst}.  The Schwarzian modes for $D$ parametrize $\diff\,S^1/\PSL(2,\R)$, but the Schwarzian modes
for $D(\alpha)$ and $T$ both parametrize $\diff\,S^1/\U(1)$.   This plus the localization procedure
implies that the one-loop determinants for $D(\alpha)$ and $T(b)$
are the same (and moreover independent of $\alpha$ and $b$).\footnote{In the localization procedure of Duistermaat and Heckman (and
Atiyah and Bott) which was exploited in \cite{SWfirst}, one considers an action of the group $\U(1)$ on a symplectic manifold
$M$, with symplectic form $\omega$ and Hamiltonian $H$.   One then considers the integral $\int_M \exp(H/\beta+\omega)$.   In general
in this situation, the one-loop determinant in expanding around a fixed point $p$ of the $\U(1)$ action depends only on the ``rotation angles'' with which the group $\U(1)$ acts on the tangent space to $M$ at $p$.   It does not depend on $H$ or $\omega$.   In our application, the relevant
$\U(1)$ group is the group of rotations of the asymptotically AdS boundary  of the two-dimensional surface.  $D(\alpha)$
and $T(b)$ have different $H$ and $\omega$, but as spaces they are the same homogeneous space 
$\diff\,S^1/\U(1)$ with the same $\U(1)$ action,
so they have the same one-loop determinants.}
We can therefore borrow the results for the determinants from eqn. (7)
of \cite{SSS} (where we set $\gamma=1/2$ as remarked in footnote \ref{normalztn}).   
So the Schwarzian path integrals for the three examples are
\begin{align} \label{schpart} Z^\Sch_D(\beta) & =\frac{\exp\left(\frac{\pi^2}{\beta}\right)}{4\pi^{1/2} \beta^{3/2}} \cr
                     Z^\Sch_{D(\alpha)}(\beta)& =\frac{\exp\left(\frac{(2\pi-\alpha)^2}{4\beta}\right)}{2\pi^{1/2} \beta^{1/2}} \cr
                     Z^\Sch_{T(b)}(\beta) & = \frac{\exp\left(-\frac{b^2}{4\beta}\right)}{2\pi^{1/2} \beta^{1/2}}.\end{align}
The JT gravity path integrals for the three examples differ from these expressions only by  an additional factor $e^{\chi\SS_0}$ coming from the Einstein-Hilbert action;
here $\chi$ is the Euler characteristic and $\SS_0$ is the ground state entropy.   Note that $\chi=1$ for $D$ and $D(\alpha)$
and $\chi=0$ for $T(b)$.

\section{First Order Correction To The Eigenvalue Distribution}\label{firstorder}

It is now straightforward to compute the first order effect of a perturbation $U(\phi)$ on the eigenvalue density $\rho(E;U)$
of a hypothetical dual matrix model.

The basic idea is that the partition function $Z_D(\beta;
U)$ of the bulk theory on a disc with asymptotically AdS boundary is interpreted as
$Z_D(\beta;U)=\int_{E_0}^\infty \d E \,e^{-\beta E}\rho(E;U)$, where $E_0$ is the threshold energy and
$\rho(E;U)$ is the density of eigenvalues.   In the present case, the disc
may contain any number of conical singularities of deficit angle $\alpha$, each coming with a factor $\veps$.   The first two terms are the
JT gravity path integrals on $D$ and $D(\alpha)$:
\begin{align}\label{dofo} Z_D(\beta;U)& = e^{\SS_0}\left(Z^\Sch_D(\beta)+\veps Z^\Sch_{D(\alpha)}(\beta)+\O(\veps^2)\right) \cr& = e^{\SS_0} \left( \frac{\exp\left(\frac{\pi^2}{\beta}\right)}{4\pi^{1/2} \beta^{3/2}}+\veps 
\frac{\exp\left(\frac{(2\pi-\alpha)^2}{4\beta}\right)}{2\pi^{1/2} \beta^{1/2}}+\O(\veps^2)\right). \end{align}
We included a factor $e^{\SS_0}$ from the Einstein-Hilbert action.    

To put $Z_D(\beta;U)$ in the form $\int_{E_0}^\infty \d E \,e^{-\beta E}\rho(E;U)$ is an exercise in Gaussian integrals.
Using
\be\label{kojo}\int_{-\infty}^\infty \d \phi\,\phi \,e^{-\beta\phi^2}\left(e^{2\pi\phi}-e^{-2\pi\phi}\right)=\frac{2\pi^{3/2} e^{\pi^2/\beta}}{\beta^{3/2}},\ee
we learn, after setting $E=\phi^2$, that
\be\label{ojo}\int_0^\infty \d E \,e^{-\beta E}\frac{\sinh 2\pi\sqrt E}{4\pi^2}=\frac{e^{\pi^2/\beta}}{4\pi^{1/2}\beta^{3/2}}.\ee
   So in JT gravity, the ground state energy is $E_0=0$ and the density of states is
\be\label{JTDens} \rho_\JT(E)=e^{\SS_0}\frac{\sinh 2\pi\sqrt E}{4\pi^2}. \ee
Similarly, 
\be\label{nojo} \int_{-\infty}^\infty \d\phi \,e^{-\beta\phi^2}\frac{e^{(2\pi-\alpha)\phi}}{2\pi}=\frac{\exp\left(\frac{(2\pi-\alpha)^2}{4\beta}\right)}{2\pi^{1/2} \beta^{1/2}}.\ee
After setting $E=\phi^2$ and including contributions both from $\phi>0$ and from $\phi<0$, we get
\be\label{tolo}\int_0^\infty\d E \,e^{-\beta E}\frac{\exp\left((2\pi-\alpha)\sqrt E\right) +\exp\left(-(2\pi-\alpha)\sqrt E)\right)}{4\pi \sqrt E}
 = \frac{\exp\left(\frac{(2\pi-\alpha)^2}{4\beta}\right)}{2\pi^{1/2} \beta^{1/2}}.\ee
 This tells us that the corrected density of states, to first order in $\veps$, is
 \be\label{gojo}\rho(E;\veps)=e^{\SS_0}\left( \frac{\sinh 2\pi\sqrt E}{4\pi^2}+\veps   \frac{\exp\left((2\pi-\alpha)\sqrt E\right) +\exp\left(-(2\pi-\alpha)\sqrt E\right)}{4\pi \sqrt E}+\O(\veps^2)    \right) . \ee

This can be written more succinctly if we remember that we started with the function $W(\phi)=2\phi+U(\phi)$, with 
$U(\phi)=2\veps e^{-\alpha\phi}$.   So to first order in $U$, we have
\be\label{nojjo} \rho(E;U)=e^{\SS_0}\left(\frac{\sinh 2\pi\sqrt E}{4\pi^2}+\frac{e^{2\pi\sqrt E}U(\sqrt E)+e^{-2\pi\sqrt E}U(-\sqrt E)}{8\pi\sqrt E}
+\O(U^2)\right).\ee

In this form, the result is certainly valid in much greater generality than the derivation.   First of all, in linear order we can simply add any
perturbations for which the derivation applies.   The derivation would apply for any function of the form
\be\label{nojko}U(\phi)=2\sum_{i=1}^r \veps_i e^{-\alpha_i\phi},~~~\pi<\alpha_i <2\pi.\ee
With this perturbation, there are $r$ different types of conical singularity, with deficit angles $\alpha_i$ and amplitudes $\veps_i$.
To linear order, we just have to add up their contributions to $\rho(E)$, and this  gives eqn. (\ref{nojjo}).
The same logic applies for 
\be\label{hoggo}U(\phi)=2\int_\pi^{2\pi} \d\alpha \,\veps(\alpha) e^{-\alpha\phi},\ee
with a large class of functions (or distributions) $\veps(\alpha)$.  
However, because eqn. (\ref{nojjo}) makes sense for any real-valued function $U$, it is natural to suspect that this answer is correct for a larger
class of functions beyond those for which the derivation actually applies.  

It is interesting to look at the high energy behavior of the  density of states $\rho(E)$  with the first order correction included.   For a simple
exponential perturbation, as in eqn. (\ref{gojo}), the condition that the high energy limit of $\rho(E)$ coincides with $\rho_\JT(E)$
is $0<\alpha<4\pi$.   This includes the range $\pi<\alpha<2\pi$ where our calculation is reliable.   The lower bound at $\alpha=0$
is reminiscent of the classical discussion in section \ref{sectwo}, where we needed $\alpha>0$ in order for $\int\d^2x \sqrt g \,\exp(-\alpha\phi)$
to make sense as a perturbation of an asymptotically $\AdS_2$ spacetime.   If we are willing to apply eqn. (\ref{nojko}) for a general
class of functions $U(\phi)$, then the condition for $\rho(E)$ to coincide with $\rho_\JT(E)$ at high energies is just
\be\label{lno} \lim_{\phi\to\infty} \frac{|U(\phi)|}{\phi}=0, \ee
assuming that $|U(\phi)|$ grows more slowly than $\exp(-4\pi \phi)$ for $\phi\to-\infty$.    If that last condition is not satisfied, matters
are more complicated, since the $e^{2\pi\sqrt E}U(\sqrt E)$ and $e^{-2\pi\sqrt E} U(-\sqrt E)$ terms in eqn. (\ref{nojko})
can both be important and there
can be a cancellation between them.

It is also instructive to expand $\rho(E)$ for small $E$.    We get
\be\label{hino}\rho(E)=e^{\SS_0}\left( \frac{E^{1/2}}{2\pi}+\frac{U(0)}{4\pi E^{1/2}} +\O(E^{3/2},UE^{1/2})\right).   \ee
On the other hand, the general behavior near threshold of a hermitian matrix model  is
$\rho(E)\sim c(E-E_0)^{1/2}$, where $E_0$ is the threshold energy and $c$ is a constant.   To reconcile this with eqn. (\ref{hino}), we have to interpret
the $U(0)E^{-1/2}$ term as coming from a shift in the threshold energy.  In other words, we expand $\rho(E)$ near threshold as
\be\label{zino}\rho(E)=\frac{e^{\SS_0}}{2\pi}\left((E-E_0)^{1/2}+\O((E-E_0)^{3/2})\right)= \frac{e^{\SS_0}}{2\pi}\left(E^{1/2}+\frac{E_0}{2 E^{1/2}}+\cdots\right), \ee
where the shifted threshold energy is
\be\label{lino} E_0=-U(0) +\O(U^2). \ee

\section{Exact Result For $\rho(E)$ When $U(0)=0$}\label{higher}

   \begin{figure}
 \begin{center}
   \includegraphics[width=2.7in]{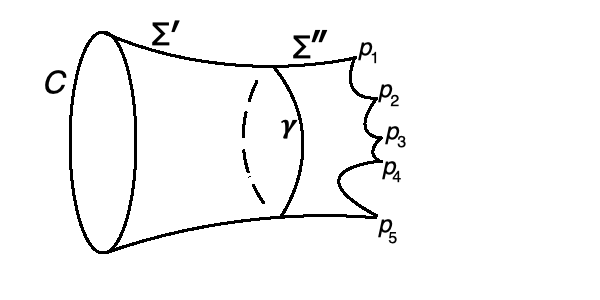}
 \end{center}
\caption{\small A disc $D$ with a number of conical singularities (five in this example) and an asymptotically AdS boundary C.
If the deficit angles are all greater than $\pi$, then by minimizing the length of a closed loop that is homologous to $C$, one
can find a closed geodesic $\gamma$ that separates $C$ from all of the conical singularities.   $D$ can thus be built by gluing 
together $\Sigma'$ and $\Sigma''$ along $\gamma$; here $\Sigma'$ is a ``trumpet'' and $\Sigma''$ is a disc with geodesic boundary
and five conical singularities.  \label{Multiple}}
\end{figure}

With what we have explained so far, it is relatively straightforward to analyze the eigenvalue density in higher orders in $U$.  
We will assume that $U$ takes the form (\ref{nojko}).  This means that a conical singularity can have any one of the deficit angles
$\alpha_i$, $i=1,\cdots,r$, with amplitude $\veps_i$.
To compute
the term of order $U^n$, for $n>1$, 
we consider a disc with asymptotically AdS boundary $C$ and $n$ conical singularities.    We have to sum over all ways of labeling
each conical singularity by one of the $\alpha_i$, and for each such choice, we have to evaluate the path integral.

If all deficit angles are greater than $\pi$,
then by minimizing the length of a closed loop that is homologous to $C$, we can find a closed geodesic $\gamma$ that separates
the asymptotically AdS boundary $C$ from all of the conical singularities.   If any deficit angle is less than $\pi$, then in general
such a $\gamma$ does not exist, as explained in the discussion of fig. \ref{FigTwo}.   The disc $D$ can be constructed
by gluing together two surfaces  $\Sigma'$ and $\Sigma''$ along $\gamma$.   Here $\Sigma'$ is a trumpet $T(b)$, where $b$
is the circumference of the loop $\gamma$.
And  $\Sigma''$
is a disc with geodesic boundary and $n$ conical singularities.\footnote{It would be possible to make a further decomposition of $\Sigma''$
by cutting on additional geodesics, but this will not be necessary for what follows.}
For given $b$, $\Sigma''$ is parametrized
by $\M_{0,b,\vec\alpha}$, the moduli space of discs (or spheres with one hole) with a geodesic boundary of length $b$,
and conical singularities with deficit angles $\vec\alpha=(\alpha_{i_1},\alpha_{i_2},\cdots \alpha_{i_n})$.   Each of the $\alpha_{i_k}$,
for $k=1,\dots,n$, corresponds to one of the exponents in the expression (\ref{nojko}) for $U$.  For each $k$, we have to sum over all choices
of $\alpha_{i_k}$, and include a weighting factor of $\veps_{i_k}$.

 We can proceed rather as in \cite{SSS}.
For fixed $b$, we multiply the path integrals on $\Sigma'$ and $\Sigma''$; then we integrate over all values of $b$, with the measure $b\,\d b$.
The path integral of $\Sigma'$ is $Z^\Sch_{T(b)}(\beta)$ from eqn. (\ref{schpart}), where $\beta$ is the renormalized length of the outer boundary 
$C$ of $\Sigma'$.   And the path integral of $\Sigma''$ is the Weil-Petersson volume of $\M_{0,b,\vec\alpha}$, multiplied by $e^{\SS_0}$ (as
$\Sigma''$ has $\chi=1$). 
So overall, the path integral of the disc $D$ with $n>1$ conical singularities is
\be\label{bogg}e^{\SS_0}\sum_{i_1,i_2,\cdots,i_n=1}^r \frac{\veps_{i_1}\veps_{i_2}\cdots \veps_{i_n}}{n!} \int_0^\infty b\,\d b \frac{\exp\left(
-\frac{b^2}{4\beta}\right)}{2\pi^{1/2}\beta^{1/2}} V_{0,b,\alpha_{i_1},\alpha_{i_2},\cdots,\alpha_{i_n}}. \ee
We will only consider small values of $n$, for which explicit formulas for the volumes are available.    For any $n$, 
 the volumes can be computed via  eqn. (\ref{nugo}), which for the case of just one hole becomes
\be\label{volform}  V_{0,b,\alpha_{i_1},\alpha_{i_2},\cdots,\alpha_{i_n}}=\int_{\bar \M_{0,1+n} }\exp\left(\omega_{0,1+n} +
\frac{b^2}{2}\psi  -\frac{1}{2}\sum_{j=1}^n (2\pi -\alpha_{i_j})^2\t\psi_j\right). \ee
This  formula can be evaluated explicitly for any $n$ using  facts explained in \cite{DW,WittenOld}, but we will only use it
to deduce some qualitative properties.

The moduli space $\M_{0,1+n}$  has real dimension  $2(n-2)$.  On the other hand, $\psi$ and the $\t\psi_j$ are
two-dimensional classes.  So $V_{0,b,\alpha_{i_1},\alpha_{i_2},\cdots,\alpha_{i_n}}$ is a polynomial in the $n$ variables
$ (2\pi -\alpha_{i_j})^2$, $j=1,\cdots,n$, of total degree at most $n-2$.    Such a polynomial is inevitably a sum of terms each of which
is independent of at least two of the variables.   Consider a term in  $V_{0,b,\alpha_{i_1},\alpha_{i_2},\cdots,\alpha_{i_n}}$ which is
independent of, for example, $\alpha_{i_k}$.     Then in the sum in eqn. (\ref{bogg}), nothing depends on ${i_k}$ except
$\veps_{i_k}$.    The sum over $i_k$ then reduces to
\be\label{sumred} \sum_{i_k=1}^r \veps_{i_k} =\frac{U(0)}{2}, \ee
where we evaluated the sum using eqn. (\ref{nojko}).    

Thus, every contribution to the path integral of a disc with $n>1$ punctures is proportional to $U(0)$.   In fact, every such contribution
is proportional to $U(0)^2$, since $V_{0,b,\alpha_{i_1},\alpha_{i_2},\cdots,\alpha_{i_n}}$ is a polynomial in which each term is
independent of at least two of the $\alpha_{i_k}$.   

Thus, when $U(0)=0$, the formula (\ref{nojjo}) for the matrix model 
density of states $\rho(E)$ is exact, at least to all orders of perturbation theory.
Any correction is proportional to $U(0)^2$.  

The condition $U(0)=0$ means, as we found in section \ref{firstorder}, that the threshold energy of the dual matrix model is
the same as in JT gravity.   When this is not the case, $\rho(E)$ can have corrections of quadratic and higher order in $U$.
In fact, there  must be such corrections, assuming that this class of perturbed
JT gravity theories really are dual to matrix models.   That is because a hermitian matrix model has the threshold behavior
$\rho\sim \sqrt{E-E_0}$ for some $E_0$, 
but in section \ref{firstorder} we found only the first two terms of an expansion of
such a function in powers of $E_0$. The higher order terms must come from contributions with $n>1$.

  When $U(0)\not=0$, there is no problem to compute higher order terms in $\rho(E)$ 
   by making eqn. (\ref{bogg}) more explicit.  We turn to this next.

\section{Perturbative Results When $U(0)\not=0$}\label{exrel}

The first nontrivial case is $n=2$.    
For any deficit angles    $\alpha_{i_1},\alpha_{i_2}$  (in the favored range\footnote{Outside this range, the moduli
space may be empty.}  $\pi<\alpha<2\pi$),
the moduli space $\M_{0,b,\alpha_{i_1},\alpha_{i_2}}$ is a point, with volume 1.
So we can perform the integral over $b$ and the sum over $i_1,i_2$  in eqn. (\ref{bogg}), to get a very simple answer
for the $n=2$ contribution to the disc partition function:
\be\label{nubb} e^{\SS_0}\frac{U(0)^2 \beta^{1/2}}{8\pi^{1/2}}. \ee

To understand this answer, it helps to go back to eqn. (\ref{dofo}), which for a general $U$ (of the form (\ref{nojko})) becomes
\be\label{wondo}  Z_D(\beta;U) = e^{\SS_0} \left( \frac{\exp\left(\frac{\pi^2}{\beta}\right)}{4\pi^{1/2} \beta^{3/2}}+\sum_{i=1}^r\veps_i
\frac{\exp\left(\frac{(2\pi-\alpha_i)^2}{4\beta}\right)}{2\pi^{1/2} \beta^{1/2}}+\O(U^2)\right).\ee
The low energy behavior is related to the behavior for large $\beta$.   
Keeping in each order of $U$ only the dominant behavior for large $\beta$, we get
\be\label{tondo} Z_D(\beta;U)=e^{\SS_0} \frac{1+U(0)\beta}{4\pi^{1/2}\beta^{3/2}} +\cdots\ee
where the omitted terms are suppressed by powers of $1/\beta$ or $U$.
Now from eqn. (\ref{nubb}), we can add the $\O(U^2)$ term and we get 
\be\label{iondo}  Z_D(\beta;U)=e^{\SS_0}\frac{1+U(0)\beta +\frac{1}{2} U(0)^2\beta^2}{4\pi^{1/2}\beta^{3/2}}+\cdots
=e^{\SS_0}\frac{\exp(\beta U(0))}{4\pi^{1/2}\beta^{3/2}}+\cdots. \ee

Since the disc partition function is interpreted in the matrix model as $\Tr\,e^{-\beta H}$, a factor $e^{\beta U(0)}$ in the
disc partition function precisely amounts to a shift in the Hamiltonian by a constant $-U(0)$.   
So in other words the result (\ref{nubb}) for the $n=2$ contribution is precisely right to represent the effect quadratic in $U(0)$
of a shift in the threshold energy by $-U(0)$.   

The expected threshold behavior (using eqn. (\ref{hino}) to determine some constants) is 
\be\label{pkol}\rho(E)\sim e^{\SS_0} \frac{\sqrt{E+U(0)}}{2\pi}=e^{\SS_0}\frac{E^{1/2}}{2\pi}\left(1+\frac{1}{2}\frac{U(0)}{E}-\frac{1}{8}\frac{U(0)^2}{E^2}+\cdots\right),\ee with a possible further shift in the threshold energy in higher orders in $U$.
We have already identified contributions that match the first two terms in this expansion, and
we  would like to interpret eqn. (\ref{nubb}) as representing an $n=2$ contribution to $\rho(E)$ that matches the third term:
\be\label{lkof} \rho_{[n=2]}(E)= -e^{\SS_0}\frac{U(0)^2}{16\pi}\frac{1}{E^{3/2}} \ee  
To justify this, we need 
\be\label{ngo}- e^{\SS_0}\frac{U(0)^2}{16\pi}\int_0^\infty \d E \,E^{-3/2}e^{-\beta E} =e^{\SS_0}\frac{U(0)^2\beta^{1/2} }{8\pi^{1/2}}.\ee
This is the $k=-1/2$ case of
\be\label{ngof} -e^{\SS_0}\frac{U(0)^2}{16\pi}\int_0^\infty \d E \,E^{k-1}e^{-\beta E} =-e^{\SS_0}\frac{U(0)^2\beta^{-k} \Gamma(k)}{16\pi},~~k>1\ee
 (note that
$\Gamma(-1/2)=-2\sqrt \pi$).  The integral (\ref{ngo}) actually does not converge, but one can define $1/E^{3/2}$ as a distribution
along the lines of the main theorem of \cite{Atiyah}, and this gives a precise meaning to (\ref{ngo}) and thence to the statement that
the $n=2$ contribution to $\rho(E)$ is as written in (\ref{lkof}).

The first case in which we have to take into account a nontrivial volume of the moduli space is $n=3$.  
The relevant volume was given in eqn. (\ref{vf}) (modulo the substitution $b\to \i(2\pi-\alpha)$):
\be\label{omigo}V_{0,b,\alpha_1,\alpha_2,\alpha_3}= 2{\pi^2}+\frac{b^2}{2}-\frac{1}{2}\sum_{i=1}^3 (2\pi-\alpha_i)^2
=-4\pi^2+\frac{b^2}{2}+\frac{1}{2}\sum_{i=1}^3(4\pi\alpha_i-\alpha_i^2). \ee   For $U$ as in eqn. (\ref{nojko}),
we have
\begin{align}\label{powery}\sum_{i=1}^r\veps_i & = \frac{U(0)}{2} \cr
                   \sum_{i=1}^r\veps_i\alpha_i & =-\frac{U'(0)}{2}\cr
                     \sum_{i=1}^r \veps_i\alpha_i^2 & = \frac{U''(0)}{2}. \end{align}
With these formulas, it is straightforward to evaluate eqn. (\ref{bogg}) and get the $n=3$ contribution to the disc partition function.
We combine this with our previous formulas and write the full disc partition function up to this order :
\begin{align}\label{fullp}Z_D(\beta;U) =& \frac{e^{\SS_0}}{4\pi^{1/2}\beta^{3/2}}\left(\exp\left(\frac{\pi^2}{\beta}\right)+2\beta \sum_{i=1}^r
\veps_i \exp\left(\frac{(2\pi-\alpha_i)^2}{4\beta}\right) + \frac{1}{2}U(0)^2 \beta^2\right. \cr &\left.+ U(0)^3\left(\frac{1}{6}\beta^3-\frac{1}{3}\pi^2\beta^2\right) +U(0)^2\left( -\frac{\pi}{2} U'(0) -\frac{1}{8} U''(0)\right)\beta^2\right)+\O(U^4). \end{align}     

Perhaps the most interesting question to ask about this formula is whether it is consistent with the expected threshold behavior
of $\rho(E)$.   To get $\rho(E)\sim (E-E_0)^{1/2}+\O((E-E_0)^{3/2})$, the disc partition function must have the form
\be\label{mustform} Z_D(\beta;U)= e^{-\beta E_0}\left(\frac{c}{\beta^{3/2}}+\O(\beta^{-5/2})\right)\ee
for some constant $c$.   In fact, eqn. (\ref{fullp}) does have this form, with
\be\label{nuttu} E_0=-U(0)+\frac{1}{2}\pi^2 U(0)^2 +\pi U(0)U'(0) +\frac{1}{4} U(0) U''(0) +\O(U^3). \ee   
So we do get the expected behavior, but with an $\O(U^2)$ correction to the threshold energy.  

We will not calculate directly the $\O(U^n)$ terms in $Z_D(\beta;U)$ with $n>3$, but the general form of these contributions  will be important in section \ref{exth}.   Taking into account the explicit factor of $\beta^{-1/2}$ in eqn.
(\ref{bogg}) and the fact that the volume that appears in that formula is a polynomial in $b^2$ of degree $n-2$, we see that the
$\O(U^n)$ contribution to $ Z_D(\beta;U)$ for $n\geq 2$ has the general form
\be\label{bubu} \frac{e^{\SS_0}}{4\pi^{1/2}\beta^{3/2}} P_n(\beta),~~~ P_n(\beta)=\sum_{k=2}^{n}a_k \beta^k. \ee
Thus $P_n(\beta)$ is a polynomial of degree $n$ whose lowest term is of order $\beta^2$. This depends in part on the fact
that the integral over $b$ always gives at least one power of $\beta$.    The polynomials $P_2(\beta)$ and
$P_3(\beta)$ can be read off from eqn. (\ref{fullp}).   

Thus if we define
\be\label{kino} G(\beta;U) = \frac{4\pi^{1/2}\beta^{3/2}}{e^{\SS_0}}   Z_D(\beta;U), \ee
then we have
\be\label{ino}G(\beta;U)=  \exp\left(\frac{\pi^2}{\beta}\right)+2\beta \sum_{i=1}^r
\veps_i \exp\left(\frac{(2\pi-\alpha_i)^2}{4\beta}\right)+\sum_{n=2}^\infty U^n P_n(\beta). \ee
Here and later, we schematically write $U^p$ to represent any term that is of order $p$ in the coefficients
$\veps_i$ of $U=\sum_i\veps_i e^{-\alpha_i\phi}$.

\section{Exact Results When $U(0)\not=0$}\label{exth}

Rather remarkably, if we assume that the threshold behavior is as expected for a hermitian matrix model, it is possible to deduce
 exact formulas for $E_0$ and for the disc partition function.  The higher order corrections that we have not explicitly calculated
 are uniquely determined by the assumption that the threshold behavior has the expected form.

  First of all, $E_0$ will have an expansion as a functional of $U$, of the form
$E_0=-U(0)+\O(U^2)$ (where the term quadratic in $U$ is actually given in eqn. (\ref{nuttu})).    In order for $ Z_D(\beta;U)$
to have the expected form $e^{-\beta E_0(U)}\beta^{-3/2}$
for large $\beta$, it must be that the function $G(\beta;U)$ (eqn. (\ref{kino})) has the following property:
  in each order in $U$, when $e^{\beta E_0(U)} G(\beta;U)$    is expanded
in a Laurent series around $\beta=\infty$, it should have only terms of nonpositive order.   
Thus, consider the expansion
\be\label{nuju}e^{\beta E_0(U)}G(\beta;U)=  \sum_{p=0}^\infty\sum_{-\infty< s\leq p}c_{p,s}  U^p \beta^s  .\ee
To get this expansion, one just makes Taylor series expansions of all the exponentials $\exp(\beta E_0(U))$,  
$\exp\left(\frac{\pi^2}{\beta}\right)$, and 
$\exp\left(\frac{(2\pi-\alpha_i)^2}{4\beta}\right)$.   
To see that the expansion takes the claimed form (with the indicated upper bound on the power of $\beta$), 
one uses the fact that $E_0\sim U$ and that the polynomials 
$P_n(\beta)$ in eqn. (\ref{bubu}) are of degree $n$.
  To get the desired large $\beta$ behavior of $ Z_D(\beta;U)$, we require that 
the coefficients $c_{p,s}$ with $s>0$ all vanish.

Most of these coefficients depend on the polynomials $P_n(\beta)$, which we have only calculated for $n=2,3$.
   But remarkably, the coefficients $c_{p,s}$ with $s=1$ only depend on the $n=0$ and $n=1$
contributions to $ Z_D(\beta;U)$.    This is true because the polynomials $P_n(\beta)$ are all divisible by
 $\beta^2$, as asserted in  eqn. (\ref{bubu}).
The upshot is that we can evaluate the condition $c_{p,1}=0$ without needing to know the contributions to $ Z_D(\beta;U)$ with $n>1$.
In other words, $c_{p,1}$ is the same as the corresponding coefficient  in the expansion of 
\be\label{zz} F(\beta;U)= \exp(\beta E_0(U)) \left(\exp\left(\frac{\pi^2}{\beta}\right)+2\beta \sum_{i=1}^r
\veps_i \exp\left(\frac{(2\pi-\alpha_i)^2}{4\beta}\right)  \right). \ee
The condition $ c_{p,1}=0$ suffices to determine $E_0(U)$ to all orders in $U$.   

The vanishing of $ c_{p,1}$ for all $p$ is equivalent to  the condition that 
\be\label{wz}\frac{1}{2\pi\i} \oint\frac{\d \beta}{\beta^2}        \exp(\beta E_0(U)) \left(\exp\left(\frac{\pi^2}{\beta}\right)+2\beta \sum_{i=1}^r
\veps_i \exp\left(\frac{(2\pi-\alpha_i)^2}{4\beta}\right)  \right)=0    \ee for all $U$,
where the contour goes once around $\beta=0$.    The integrals are modified Bessel functions, so we get a closed form
equation for $E_0(U)$:
\be\label{kindo} \frac{\sqrt{E_0}}{\pi}I_1(2\pi \sqrt{E_0}) +2\sum_{i=1}^r \veps_i I_0((2\pi-\alpha_i)\sqrt{E_0}) =0. \ee
This transcendental equation has to be supplemented with the condition that $E_0(U)=-U(0)+\O(U^2)$.
Using $I_1(z)=\frac{z}{2}+\frac{z^3}{16}+\O(z^5)$, $I_0(z)=1+\frac{z^2}{4}+\O(z^4)$, along with eqn. (\ref{powery}), one
can recover the quadratic term as presented in eqn. (\ref{nuttu}).

Eqn. (\ref{kindo})  determines the {\it exact} threshold energy, for $U$ of the form that we have assumed, at least to all orders of 
perturbation theory, assuming that  the model can be described as a hermitian matrix model.     
Once $E_0(U)$ is known, the contributions of order $U^n$ to the disc
partition function are  uniquely determined by the vanishing of the coefficients $c_{p,s}$ in eqn. (\ref{nuju}) with $s>1$.   We have
only directly verified  that these contributions have the expected form (and thus that the threshold behavior is as expected)
for $n=2$ and $n=3$.

   It is actually possible to determine the exact form of the disc partition function,
assuming that the model is a hermitian matrix model.   Consider the function $F(\beta;U)$ as a function of $\beta$ for fixed $U$.
It has essential singularities at $\beta=0$ and $\beta=\infty$ and otherwise is holomorphic in $\beta$.   
Such a function can
be decomposed as $F(\beta;U)=F_0(\beta;U)-F_\infty(\beta;U)$, where $F_0(\beta;U)$ is a holomorphic throughout the complex
$\beta$ plane and at infinity, with only an essential singularity at $\beta=0$, and $F_\infty(\beta;U)$ is an entire function with only
an essential singularity at infinity.   The decomposition is unique if we stipulate that $F_\infty(0;U)=0$.   In the present case, assuming
that the disc partition function has the expected threshold behavior, the
decomposition is very simple:  $F_0(\beta;U)=e^{\beta E_0(U)} G(\beta;U)$ and $F_\infty(\beta;U)=-e^{\beta E_0(U)}\sum_{n=2}^\infty U^n P_n(\beta)$.  (Since $P_n(0)=0$ for all $n$, this is consistent with $F_\infty(0;U)=0$.)
So $G(\beta;U)=e^{-\beta E_0(U)} F_0(\beta;U)$, and this, of course, determines the disc partition function $ Z_D(\beta;U)$ via eqn.
(\ref{kino}).

Explicitly, the decomposition $F(\beta;U)=F_0(\beta;U)-F_\infty(\beta;U)$ can be made as follows.   For a function $F(\beta;U)$ that is 
holomorphic except
at 0 and $\infty$, we can write
\be\label{nuccu} F(\beta;U)=\frac{1}{2\pi \i} \left(\oint_{C_1} \d w F(w;U) \frac{\beta}{w(w-\beta)} - \oint_{C_2} \d w F(w;U) \frac{\beta}{w(w-\beta)}  \right),\ee
where $C_1$ is a circle centered at $w=0$ with radius greater than $|\beta|$, and $C_2$ is a circle centered at $w=0$ with radius less than
$|\beta|$.   Here $\oint_{C_1} \d w F(w;U) \frac{\beta}{w(w-\beta)}$ is holomorphic in $\beta$ except at $\beta=\infty$, and vanishes
at $\beta=0$; $\oint_{C_2} \d w F(w;U) \frac{\beta}{w(w-\beta)}$ is holomorphic in $\beta$ except at $\beta=0$, and approaches
a constant for $\beta\to\infty$.   So $F_0(\beta;U)=-\frac{1}{2\pi \i}\oint_{C_2} \d w F(w;U) \frac{\beta}{w(w-\beta)}$, $F_\infty(\beta;U)
=-\frac{1}{2\pi \i}\oint_{C_1} \d w F(w;U) \frac{\beta}{w(w-\beta)}$.    Thus finally 
\be\label{inz} G(\beta;U)=\frac{e^{-\beta E_0(U)} }{2\pi\i}\oint_{C_2} \d w F(w;U) \frac{\beta}{w(\beta-w)}.\ee

The disc partition function is therefore
\be\label{polyo} Z_D(\beta;U)=\frac{e^{\SS_0}}{4\pi^{1/2}\beta^{3/2}}G =\frac{e^{\SS_0}}{4\pi^{1/2}}\frac{1}{2\pi\i}\oint_{C_2}\frac{\d w}{w}
F(w;U)\frac{e^{-\beta E_0(U)}}{\beta^{1/2}(\beta-w)}.\ee
This can be turned into a formula for the density of states by using 
\be\label{invlt} \int_{E_0}^\infty \d E \frac{1}{\sqrt w}e^{-\beta E+w(E-E_0)}  \Erf(\sqrt{w(E-E_0)}) =\frac{e^{-\beta E_0}}{\beta^{1/2}(\beta-w)},\ee
where $\Erf$ is the error function
\be\label{ninvit} \Erf(x)=\frac{2}{\sqrt\pi} \int_0^x\d t\, e^{-t^2}=\frac{2x}{\sqrt\pi}\int_0^1\d s \,e^{-x^2 s^2}. \ee
So
\be\label{homigo}Z_D(\beta;U)=\int_{E_0}^\infty \d E \,\rho(E;U) e^{-\beta E}\ee
with 
\be\label{nomigo}\rho(E;U)=\frac{e^{\SS_0}\sqrt{E-E_0(U)}}{2\pi}\int_0^1\d s\frac{1}{2\pi \i}\oint_{C_2}\frac{\d w}{w} F(w;U)\exp(w(E-E_0)(1-s^2)).\ee
Using the explicit form (\ref{zz}) of the function $F(w;U)$, the integral over $w$ can be expressed in terms of modified Bessel
functions, rather as before.  Finally
\begin{align}\label{toldox}\rho(E;U)=&\frac{e^{\SS_0}\sqrt{E-E_0(U)} } {2\pi} \int_0^1\d s \biggl( I_0(2\pi f(s)) 
+\sum_{i=1}^r \veps_i\frac{2\pi-\alpha_i}{f(s)} I_1((2\pi-\alpha_i)f(s)) \biggr), \end{align}
with
\be\label{watt} f(s)=((1-s^2) E+s^2 E_0)^{1/2}.\ee

\section{Higher Order Correlators}\label{hco}

So far we have computed contributions to the disc partition function with various numbers of conical singularities. These contributions
provide information about the eigenvalue density $\rho(E)$ in a deformation of JT gravity.   Since this eigenvalue density is {\it a priori}
unknown, much of what we learn does not really test the hypothesis that the deformed theory is a matrix model.   We did get an interesting
test of this hypothesis by requiring that the theshold behavior of $\rho(E)$ should have the square root behavior expected in a hermitian
matrix model.     

If it is true that the theory under study is such a matrix model, then once $\rho(E)$ is known, the path integral on an oriented two-manifold
of any other topology is uniquely determined.   Thus we can test the hypothesis that the deformed JT gravity theory is a matrix model
just by computing for some other topology and comparing to expectations for a hermitian matrix model with the eigenvalue density inferred
from the disc.   We will pursue this program far enough to make a number of interesting checks.

We will consider a Riemann surface $\Sigma$ 
of genus 0 with $p>1$ holes, each with an asymptotically  AdS boundary.   In terms of the dual quantum mechanical
system, this means that we will compute the genus 0 contribution to the connected correlation function of a product of $p$ partition functions,
\be\label{wollo}\biggl\langle\Tr\,\exp(-\beta_1H)\,\Tr\,\exp(-\beta_2 H)\cdots \Tr\,\exp(-\beta_p H) \biggr\rangle_c .\ee 
Here $\beta_j$, for $j=1,\cdots,p$,  is the renormalized circumference of the $j^{th}$ asymptotically AdS boundary.  

   \begin{figure}
 \begin{center}
   \includegraphics[width=2.7in]{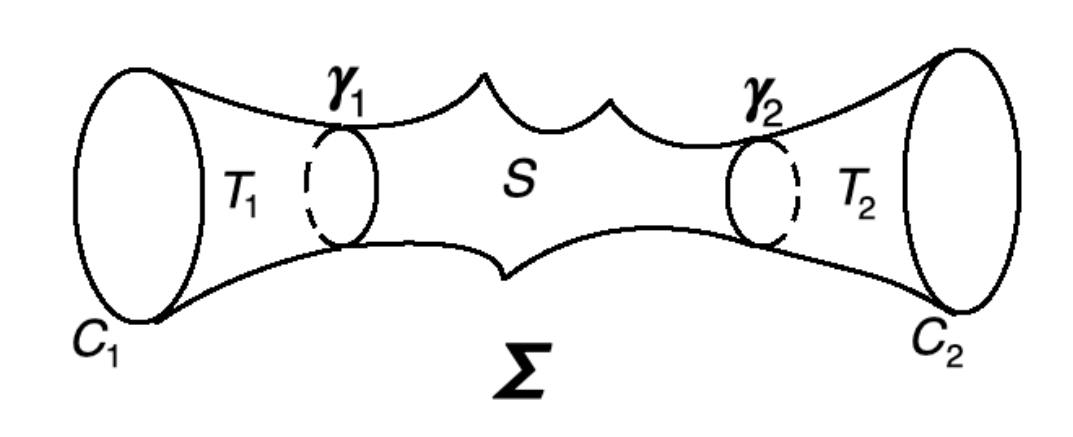}
 \end{center}
\caption{\small Pictured is a $p$-holed sphere $\Sigma$ with $p$ asymptotically AdS boundaries and $n$ conical singularities (drawn here
for $p=2$, $n=3$).   Assuming that all deficit  angles are in the favored range $\pi<\alpha<2\pi$,    $\Sigma$ can be usefully
decomposed by finding, for each asymptotically AdS boundary $C_i$, a geodesic $\gamma_i$ that is homologous to it.  Cutting
on the $\gamma_i$ decomposes $\Sigma$ into a union of $p$ trumpets $T_i$, one for each $C_i$, and a central portion $S$ that
is a sphere with $p$ geodesic boundaries and $n$ conical singularities.      \label{ComplexP}}
\end{figure}

The strategy of the computation is very similar to what was done in \cite{SSS}.    For each asymptotically AdS boundary $C_i$,
one finds a closed geodesic $\gamma_i$ that is homologous to $C_i$.   The $\gamma_i$ exist and are unique assuming the
deficit angles are in the usual range $\pi<\alpha_i<2\pi$.   Cutting on the $\gamma_i$ decomposes $\Sigma$ into a union of $p$
trumpets $T_i$, one for each $C_i$, together with a central portion $S$.  $S$ is a $p$-holed sphere with geodesic boundaries $\gamma_i$
and $n$ conical singularities.  See fig. \ref{ComplexP}.
 Let $b_i$ be the circumference of $\gamma_i$.   The path integral on $\Sigma$ is computed
by multiplying the path integrals on $T_i$ and $S$ and integrating over the $b_i$ with the familiar measure $\prod_i b_i\d b_i$.
The path integral on a trumpet $T_i(b_i)$ was given in eqn. (\ref{schpart}), and the path integral on $S$ when $\vec \alpha$ is 
specified  is  $e^{(2-p)\SS_0}$ times
 the Weil-Petersson
volume $V_{0,\vec b,\vec\alpha}$, where $\vec b=(b_1,b_2,\cdots,b_p)$ and $\vec\alpha=(\alpha_1,\alpha_2,\cdots,\alpha_n)$.  
To get the path integral in a deformation of JT gravity with $U(\phi)=\sum_i\veps_i e^{-\alpha_i\phi}$, as usual we have to 
sum 
over all choices $\vec\alpha=(\alpha_{i_1},\alpha_{i_2},\cdots \alpha_{i_n})$, with weight $\veps_{i_1}\veps_{i_2}\cdots\veps_{i_n}$, and
divide by $n!$. 
So finally  the path integral on $\Sigma$ is
\be\label{indigo} Z_\Sigma=e^{\SS_0(2-p)}\frac{1}{n!}\sum_{i_1i_2\cdots i_n=1}^r \veps_{i_1}\veps_{i_2}\cdots \veps_{i_r}\prod_{j=1}^p \int_0^\infty b_j\,\d b_j \frac{\exp\left(-\frac{b_j^2}{4\beta_j}\right)}{2\pi^{1/2} \beta_j^{1/2}}
~V_{0,\vec b,\vec\alpha}. \ee

First we consider the case $p=2$.   In JT gravity, the eigenvalue distribution $\rho(E)$  of the dual matrix model is supported on a half-line
$[0,\infty)$, and in the deformations of JT gravity that we are considering in this paper, it is supported on a half-line $[E_0,\infty)$,
with $E_0$ in general non-zero.    A remarkable fact about hermitian matrix models with $\rho(E)$ supported on a half-line is
  that in lowest order in $e^{\SS_0}$
-- that is in genus zero -- 
the connected expectation value
$\bigl\langle \Tr\,\exp(-\beta_1H)\,\Tr\,\exp(-\beta_2H)\bigr\rangle_c$  depends only on the threshold energy $E_0$
and is otherwise universal.\footnote{This can be proved using ``topological recursion'' \cite{EO}.   See for example
section 4.1.2 of \cite{SW}.}    Let us see how this result is reproduced in models obtained by deforming JT gravity.

\begin{figure}
 \begin{center}
   \includegraphics[width=2.7in]{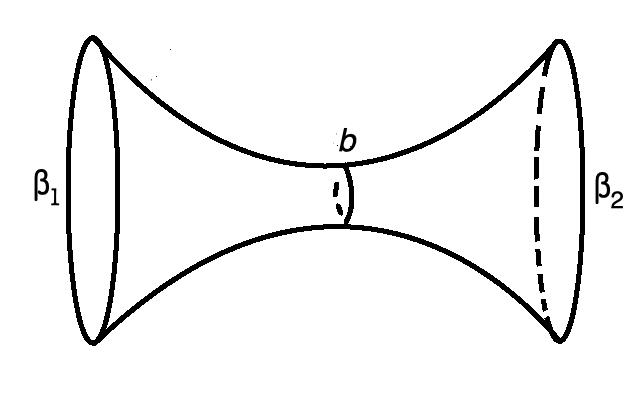}
 \end{center}
\caption{\small   A ``double trumpet.''   Two trumpets are glued together along a closed geodesic  of circumference $b$.
The outer boundaries of the two trumpets have renormalized lengths $\beta_1$ and $\beta_2$.     \label{DoubleTrumpet}}
\end{figure}

First let us recall how the connected two-point function of the matrix trace was computed in \cite{SSS}.   The relevant two-manifold
is a ``double trumpet,'' obtained by gluing together two trumpets along a closed geodesic of circumference $b$ (fig. \ref{DoubleTrumpet}).   
To compute 
$\bigl\langle \Tr\,\exp(-\beta_1H)\,\Tr\,\exp(-\beta_2H)\bigr\rangle_c$, one takes the outer boundaries of the two trumpets to have
renormalized lengths $\beta_1$ and $\beta_2$.   Note that in this particular case, with no conical singularities at all, the central
region $S$ of fig. \ref{ComplexP} is absent.   As soon as one or more conical singularities are present, this region will appear.

To compute the path integral of the double trumpet, one simply takes the product of the path integrals of the two trumpets -- which are given in 
eqn. (\ref{schpart}) -- and integrates over $b$ with the usual measure $b\,\d b$.  The result is 
\be\label{mixt}\int_0^\infty b\,\d b \frac{\exp\left(-\frac{b^2}{4\beta_1}\right)\exp\left(-\frac{b^2}{4\beta_2}\right)}{4\pi \beta_1^{1/2}\beta_2^{1/2}}
=\frac{\beta_1^{1/2}\beta_2^{1/2}}{2\pi(\beta_1+\beta_2)}. \ee
This agrees with the universal matrix model answer for the case that $E_0=0$.
If we simply add a constant $E_0$ to the matrix Hamiltonian, this multiplies $\Tr\,\exp(-\beta H)$ by $\exp(-\beta E_0)$, and hence
the connected two-point function of the matrix trace would become
\be\label{nixt}\biggl\langle \Tr\,\exp(-\beta_1H)\,\Tr\,\exp(-\beta_2H)\biggr\rangle_c = e^{-(\beta_1+\beta_2)E_0}
 \frac{\beta_1^{1/2}\beta_2^{1/2}}{2\pi(\beta_1+\beta_2)}. \ee
which is the more general matrix model answer.
So if it is true that the deformations of JT gravity studied in the present paper are dual to matrix models, those models have
to reproduce this answer, with the threshold energy $E_0(U)$ that was calculated in sections 
\ref{exrel} and \ref{exth}.  

To explore this point, let us analyze the contribution of order $U^n$ to this correlation function.   Let $b_1$ and $b_2$ be the 
circumferences of the geodesics $\gamma_1$ and $\gamma_2$ of fig. \ref{ComplexP}.    Eqn. (\ref{indigo}) gives in this case
\be\label{nogg} \sum_{i_1,i_2,\cdots,i_n=1}^r \frac{\veps_{i_1}\veps_{i_2}\cdots \veps_{i_n}}{n!} \int_0^\infty b_1\d b_1 \frac{\exp\left(
-\frac{b_1^2}{4\beta_1}\right)}{2\pi^{1/2}\beta_1^{1/2}}  \int_0^\infty b_2\d b_2 \frac{\exp\left(
-\frac{b_2^2}{4\beta_2}\right)}{2\pi^{1/2}\beta_2^{1/2}}V_{0,b_1,b_2,\alpha_{i_1},\alpha_{i_2},\cdots,\alpha_{i_n}}. \ee
Now we can make an argument that will be familiar from section \ref{higher}.    $V_{0,b_1,b_2,\alpha_{i_1},\alpha_{i_2},\cdots,\alpha_{i_n}}$
is the volume of the moduli space $\M_{0,b_1,b_2,\alpha_{i_1},\alpha_{i_2},\cdots,\alpha_{i_n}}$ that parametrizes a sphere with
two holes with geodesic boundaries of specified lengths  and $n$ conical singularities.   
From eqn. (\ref{nugo}), the volume of this moduli space is
\be\label{longvol}  V_{0,b_1,b_2,\alpha_{i_1},\alpha_{i_2},\cdots,\alpha_{i_n}}=\int_{\bar \M_{0,2+n} }\exp\left(\omega_{0,2+n} +
\frac{b_1^2}{2}\psi_1+\frac{b_2^2}{2}\psi_2  -\frac{1}{2}\sum_{j=1}^n (2\pi -\alpha_{i_j})^2\t\psi_j\right). \ee
The real dimension of this moduli space is
$2n-2$.   The $\t\psi_j$ are two-dimensional cohomology classes.   So $ V_{0,b_1,b_2,\alpha_{i_1},\alpha_{i_2},\cdots,\alpha_{i_n}}$ is
a sum of terms each of which is independent of one of the $n$ variables $\alpha_{i_j}$, for some $j$.     Upon inputting this fact in eqn. (\ref{nogg}), one learns,
exactly as in the discussion of eqn. (\ref{bogg}), that the path integral for any $n$ is proportional to $U(0)$.    

Therefore, if $U(0)=0$, all corrections to the double trumpet vanish, and the JT result for 
$\langle \Tr\,\exp(-\beta_1H)\,\Tr\,\exp(-\beta_2H)\rangle_c$ remains valid to all orders in $U$.   
But this is what we would expect from random matrix theory.    If $U(0)=0$, then $E_0=0$
and the universality of this particular correlation function in random matrix theory implies that we should expect no correction to the JT
gravity result.

What happens if $U(0)\not=0$ and therefore $E_0\not=0$?  The first case is $n=1$.  For $n=1$, the relevant moduli space
is a point, so its volume is 1.     So eqn. (\ref{nogg}) reduces to
\be\label{logg}\sum_{i=1}^r\veps_i \int_0^\infty b_1\d b_1 \frac{\exp\left(
-\frac{b_1^2}{4\beta_1}\right)}{2\pi^{1/2}\beta_1^{1/2}}  \int_0^\infty b_2\d b_2 \frac{\exp\left(
-\frac{b_2^2}{4\beta_2}\right)}{2\pi^{1/2}\beta_2^{1/2}}=\frac{U(0)\beta_1^{1/2} \beta_2^{1/2}}{2\pi}.\ee
Since $E_0=-U(0)+\O(U^2)$, this is in agreement with the $\O(U)$ term in the expected result (\ref{nixt}).

For $n=2$, eqn. (\ref{nogg}) becomes
\be\label{fogg}     \frac{1}{2}\sum_{i_1,i_2=1}^r \veps_{i_1}\veps_{i_2}  \int_0^\infty b_1\d b_1 \frac{\exp\left(
-\frac{b_1^2}{4\beta_1}\right)}{2\pi^{1/2}\beta_1^{1/2}}  \int_0^\infty b_2\d b_2 \frac{\exp\left(
-\frac{b_2^2}{4\beta_2}\right)}{2\pi^{1/2}\beta_2^{1/2}}V_{0,b_1,b_2,\alpha_{i_1},\alpha_{i_2}}. \ee
From eqn. (\ref{vf}), with the usual substitution $b\to \i(2\pi-\alpha)$, the volume is now
\be\label{nhb}V_{0,b_1,b_2,\alpha_1,\alpha_2}
=-2\pi^2 +\frac{1}{2}\sum_{i=1,2}b_i^2+\frac{1}{2} \sum_{j=1,2}(4\pi \alpha_j-\alpha_j^2).\ee  
Using eqn. (\ref{powery}) and following similar steps to the previous derivations,
we evaluate the sums and integrals in eqn. (\ref{fogg}) and get
\be\label{winzo}\frac{\beta_1^{1/2}\beta_2^{1/2}}{2\pi}\left(\frac{1}{2}U(0)^2(\beta_1+\beta_2) -\frac{\pi^2}{2}U(0)^2 -\pi U(0)U'(0)
-\frac{U(0)U''(0)}{4}\right). \ee
This agrees with the matrix model prediction (\ref{nixt}) for the term of order $U^2$ in this correlator, given the formula (\ref{nuttu}) for $E_0(U)$.

   \begin{figure}
 \begin{center}
   \includegraphics[width=2.7in]{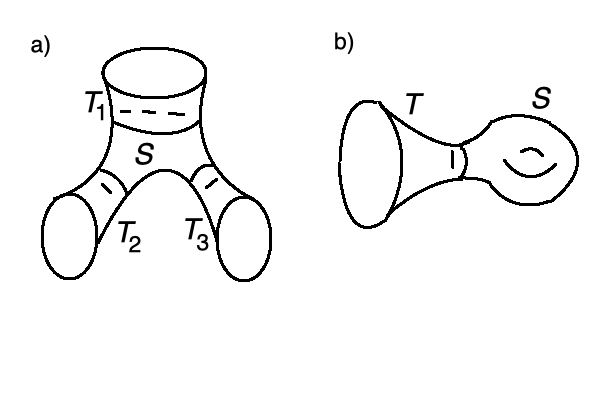}
 \end{center}
\caption{\small  (a) A three-holed sphere with three asymptotically AdS boundaries is built by gluing three trumpets $T_i$
onto a three-holed sphere $S$ with geodesic boundaries.  (b)   A one-holed torus with asymptotically AdS boundary is built
by gluing a trumpet $T$ onto a one-holed torus $S$ with geodesic boundary.   In both (a) and (b), conical singularities can be added
to $S$.     \label{LastEx}}
\end{figure}

Now we will move on to consider the expectation value of a product of three traces.   Suppose that the expansion of
the density of states near threshold is
\be\label{nubz} \rho(E)=e^{\SS_0}\left(x (E-E_0)^{1/2} +y(E-E_0)^{3/2}+\O(E-E_0)^{5/2}\right)\ee
with constants $x,y$.   All formulas will  be invariant under a common scaling $e^{\SS_0}\to t e^{\SS_0}$,
$x\to t^{-1} x$ , $y\to t^{-1}y$.   
In deformed JT gravity, one has\footnote{It is straightforward to get these formulas by expanding eqn.
(\ref{nojjo}) if one assumes that $U(0)=0$.   To verify that $x$ and $y$ do not depend on $U(0)$ (in linear order in $U$) 
is more straightforward from
eqn. (\ref{toldox}), where the shift in the threshold energy has been explicitly factored out.}
\begin{align}\label{zombo} x & = \frac{1}{2\pi}+\frac{1}{8\pi}\left(4\pi U'(0)+U''(0)\right) +\O(U^2)\cr
             y & = \frac{\pi}{3} +\frac{1}{8\pi}\left(\frac{8\pi^3}{3}U'(0) +2\pi^2 U''(0) +\frac{2\pi}{3}U'''(0) +\frac{1}{12}U''''(0)\right)+\O(U^2),\end{align}
             and of course $E_0=-U(0)+\O(U^2)$.

The matrix model prediction for the genus zero contribution to  the connected correlator of a product of three partition functions is 
\be\label{wubz}\biggl\langle \prod_{i=1}^3\Tr\,\exp(-\beta_i H)\biggr\rangle_c = \frac{e^{-\SS_0}}{x} \frac{1}{2\pi^{5/2}}\exp\left(-\sum_{i=1}^3\beta_i E_0\right)
\prod_{i=1}^3 \beta_i^{1/2}. \ee
To deduce this result, one starts with the matrix model prediction for the genus 0 contribution to the product of three resolvents
\be\label{waggo}\biggl\langle \prod_{i=1}^3\Tr\frac{1}{x_i-H}\biggr\rangle_c.  \ee
This can be obtained via topological recursion; for example, see section 4.1.3 of \cite{SW}.   Then, via an inverse Laplace transform,
one  arrives at eqn. (\ref{wubz}).

In JT gravity, the genus zero contribution to the connected correlator in eqn. (\ref{wubz}) is computed by a path integral
on a three-holed sphere $\Sigma$ 
with three asymptotically AdS boundaries with renormalized circumferences $\beta_i$.  Proceeding as in 
\cite{SSS}, this manifold is built by gluing three trumpets $T_i$ onto a three-holed sphere $S$ with geodesic boundaries (fig. \ref{LastEx}(a)).
The moduli space of $S$ is a point, with volume 1.   So the path integral on $\Sigma$ is just the product of the path integrals
on the three trumpets, times $e^{-\SS_0}$:
\be\label{zaggo}e^{-\SS_0}
 \prod_{i=1}^3\int_0^\infty b_i\d b_i \frac{e^{-b_i^2/4\beta_i}}{2\pi^{1/2}\beta_i^{1/2}}=e^{-\SS_0}\frac{\beta_1^{1/2}\beta_2^{1/2}\beta_3^{1/2}}{\pi^{3/2}}. \ee
This formula is in accord with eqn. (\ref{wubz}), since $E_0=0$ and $x=1/2\pi$ in JT gravity.   

In deforming away from JT gravity, we will only work to first order in $U$.   Thus, we have to evaluate the path integral with a single conical
singularity added to the three-holed sphere $S$ in fig. \ref{LastEx}(a).    The resulting moduli space $\M_{0,\vec b,\alpha}$ has volume
\be\label{aggo}V_{0,\vec b,\alpha}=\frac{1}{2}\sum_{i=1}^3 b_i^2+2\pi\alpha-\frac{1}{2}\alpha^2.\ee  This
follows from eqn. (\ref{vf}) with the usual substitution $b\to \i (2\pi-\alpha)$.   For $U(\phi)=\sum_{i=1}^r \veps_i e^{-\alpha_i\phi}$, it
is convenient to first sum over all choices of the deficit angle:
\be\label{baggo}\sum_{i=1}^r\veps_i V(0,\vec b,\alpha_i)=\frac{1}{4}U(0)\sum_{i=1}^3 b_i^2-\pi U'(0)-\frac{1}{4}U''(0). \ee
The path integral is then
\begin{align}\label{saggo}e^{-\SS_0}
 \prod_{i=1}^3\int_0^\infty b_i\d b_i & \frac{e^{-b_i^2/4\beta_i}}{2\pi^{1/2}\beta_i^{1/2}}\left( \frac{1}{4}U(0)\sum_{j=1}^3 b_j^2-\pi U'(0)-\frac{1}{4}U''(0)\right) \cr &=\frac{\beta_1^{1/2}\beta_2^{1/2}\beta_3^{1/2}}{\pi^{3/2}}\left(U(0)\sum_{j=1}^3\beta_j -\pi U'(0)-\frac{1}{4}U''(0)\right).
 \end{align}   This agrees with the expected $\O(U)$ contribution in eqn. (\ref{wubz}), given the formula (\ref{zombo}) for $x$
 along with  $E_0=-U(0)+\O(U^2)$.
 
 As a final example, we will consider the genus 1 correction to the partition function, which we will write as $\biggl\langle \Tr\,\exp(-\beta H)\biggr
 \rangle_1$.   First of all, the matrix model prediction is
 \be\label{kinno} \biggl\langle \Tr\,\exp(-\beta H)\biggr
 \rangle_1=e^{-\SS_0}\left(\frac{y}{x^2} \frac{\beta^{1/2}}{16\pi^{3/2}}+\frac{1}{x}\frac{\beta^{3/2}}{24 \pi^{3/2}}  \right)e^{-\beta E_0} . \ee
 To get this formula, one can first use topological recursion to obtain the genus one contribution to the expectation value of the trace of the
 resolvent $\langle \Tr \frac{1}{x-H}\rangle_1$; for example, see section 4.1.4 of \cite{SW}.   Then an inverse Laplace transform leads to
 eqn. (\ref{kinno}).   
 
 To compute $ \biggl\langle \Tr\,\exp(-\beta H)\biggr
 \rangle_1$ in JT gravity, we need to do a path integral on a one-holed torus $\Sigma$ with asymptotically AdS boundary.   This manifold
 can be constructed by gluing a trumpet $T$ to a one-holed torus $S$ with geodesic boundary (fig.
 \ref{LastEx}(b)).   The path integral on $S$ is actually a rather subtle example, because a one-holed torus with geodesic boundary,
 like a torus with one puncture (to which it reduces when the circumference $b$ of the boundary goes to zero) has a $\Z_2$ group
 of automorphisms.   One can either say that the volume of the moduli space $\M_{1,b}$ is $\frac{1}{24}(4\pi^2+b^2)$, which is the
 formula in \cite{M}, and that the path integral on $S$ is $1/2$ of this, to account for the symmetry, or one can include
 the factor of $1/2$ in the definition of the volume  (this is analogous to a common procedure
  in algebraic geometry).    Either way, the path integral on $S$ in JT gravity
 is $\frac{1}{48}(4\pi^2+b^2)$.    The path integral on $\Sigma$ is
 obtained as usual by multiplying the path integrals on $T$ and $S$ and integrating over $b$:
 \be\label{inno}e^{-\SS_0}\int_0^\infty b\,\d b \frac{\exp\left(-\frac{b^2}{4\beta}\right)}{2\pi^{1/2}\beta^{1/2}} \frac{1}{48}(4\pi^2+b^2)
 =e^{-\SS_0}\frac{1}{12\pi^{1/2}\beta^{1/2}}(\pi^2\beta+\beta^2). \ee
 Using the values $E_0=0$, $x=1/2\pi$, $y=\pi/3$ in JT gravity, one finds that this is in accord with (\ref{kinno}).
 
 In deforming away from JT gravity, we will work only to first order in $U$.  This means that we add one conical singularity to $S$.
 The corresponding volume is, after a substitution $b\to \i(2\pi-\alpha)$ of a formula in  \cite{M},
 \be\label{yto} V_{1,b,\alpha}=\frac{1}{192}\left(b^2(8\pi^2+b^2)+(4\pi \alpha-\alpha^2)(2b^2+8\pi^2)+(4\pi\alpha-\alpha^2)^2\right). \ee
Hence
\begin{align}\label{pyto}\sum_{i=1}^r\veps_i V_{1,b,\alpha_i}=\frac{1}{384} &\left(  U(0) b^2(8\pi^2+b^2)+U''''(0)+8\pi U'''(0) \right.\cr &\left.
+8\pi^2 U''(0) -32 \pi^3 U'(0) -2 b^2(4\pi U'(0)+U''(0))
\right) .\end{align}
So the path integral on $\Sigma$ is
\begin{align}
\label{plyto} e^{-\SS_0}\int_0^\infty& b\,\d b \frac{\exp\left(-\frac{b^2}{4\beta}\right)}{2\pi^{1/2}\beta^{1/2}} 
\sum_{i=1}^r \veps_i V_{1,b,\alpha_i}=e^{-\SS_0}\frac{1}{12\pi^{1/2}\beta^{1/2}}\biggl( U(0)(\pi^2\beta^2+\beta^3)\biggr.
\cr & \biggl. +\frac{\beta}{32}\bigl(U''''(0) +8\pi U'''(0) +8\pi^2 U''(0) -32\pi^3 U'(0) -8\beta (4\pi U'(0)+U''(0))   \bigr)\biggr)  .\end{align}
This is in accord with (\ref{kinno}), as one finds after using the formulas that express $x,y, $ and $E_0$ to first order in $U$.

We will leave the subject here, with one last remark.   As one goes to higher orders in the expansion in $e^{-\SS_0}$, topological
recursion of the matrix model gives formulas that involve higher and higher derivatives of $\rho(E)$ at threshold.    Qualitatively, this matches
the fact that on the gravity side, in higher orders in this expansion, one meets moduli spaces whose volumes are polynomials in the deficit angles
of increasing degree, leading to formulas that involve higher and higher derivatives of $U$ at threshold.   Perhaps a general proof that the
two expansions agree to all orders can be constructed by adapting the techniques used by Mirzakhani \cite{M} to compute
volumes of moduli spaces.

\noindent{\it Acknowledgment}   Research  supported in part by  NSF Grant PHY-1911298.

\bibliographystyle{unsrt}

\end{document}